\documentclass[11pt]{article}

\usepackage[T1]{fontenc}
\usepackage[utf8]{inputenc}
\usepackage[a4paper,margin=1in]{geometry}

\usepackage{graphicx}
\usepackage{booktabs}
\usepackage{array}
\usepackage{caption}
\usepackage{amsmath}
\usepackage{xcolor}
\usepackage{hyperref}
\usepackage{microtype}
\usepackage{tabularx}
\usepackage{makecell}

\usepackage{algorithm}
\usepackage{algpseudocode}

\usepackage{fvextra}
\usepackage{placeins}
\usepackage{comment}
\usepackage{environ}

\DefineVerbatimEnvironment{Code}{Verbatim}{
  breaklines,
  breaksymbol={},
  frame=single,
  fontsize=\small
}

\DefineVerbatimEnvironment{PromptCode}{Verbatim}{
  breaklines=true,
  breakanywhere=true,
  fontsize=\small,
  frame=single,
  framesep=2mm,
  rulecolor=\color{black}
}

\usepackage[numbers]{natbib}
\date{}

\newcommand{\Description}[1]{}
\excludecomment{CCSXML}
\newcommand{\ccsdesc}[2][]{ }

\definecolor{commentBlue}{HTML}{4285F4}
\definecolor{commentRed}{HTML}{EA4335}
\definecolor{commentGreen}{HTML}{34A853}
\definecolor{commentPurple}{HTML}{9C27B0}
\definecolor{commentOrange}{HTML}{FF9800}

\makeatletter
\newif\ifCommentScopeFirstAuthor
\CommentScopeFirstAuthortrue
\renewcommand{\author}[1]{%
  \ifCommentScopeFirstAuthor
    \gdef\@author{#1}%
    \CommentScopeFirstAuthorfalse
  \else
    \g@addto@macro\@author{\quad #1}%
  \fi}
\newcommand{\CommentScopeAffiliation}{}

\let\CommentScopeArticleMakeTitle\maketitle
\let\CommentScopeOriginalAbstract\abstract
\let\CommentScopeOriginalEndAbstract\endabstract
\newcommand{\CommentScopeAbstractBody}{}
\RenewEnviron{abstract}{\global\let\CommentScopeAbstractBody\BODY}
\newcommand{\CommentScopeKeywords}{}
\newcommand{\keywords}[1]{}
\renewcommand{\keywords}[1]{\gdef\CommentScopeKeywords{#1}}
\newcommand{\CommentScopeWordCount}{}
\newcommand{\wordcount}[1]{\gdef\CommentScopeWordCount{#1}}
\renewcommand{\maketitle}{%
  \g@addto@macro\@author{\\[6pt]\small \CommentScopeAffiliation}%
  \CommentScopeArticleMakeTitle
  \CommentScopeOriginalAbstract
  \CommentScopeAbstractBody
  \CommentScopeOriginalEndAbstract
  \par\noindent\textbf{Keywords:} \CommentScopeKeywords\par
  \par\noindent\textbf{Word count:} \CommentScopeWordCount\par
  \vspace{1em}}
\makeatother

\begin{document}

\title{Embedding User Comments into Long-Form Reading: Design and Evaluation of Contextual Comment Presentation}


\author{
Shuai Chen\textsuperscript{*} \quad
Lei Han \quad
Zhaoman Zhong
\\[6pt]
\small Jiangsu Ocean University,
Lianyungang, Jiangsu, China
\\[3pt]
\small
\textsuperscript{*}Corresponding author: \texttt{shuai.chen@jou.edu.cn}
}


\begin{abstract}

Online comments offer perspectives on long-form articles but are often separated from the passages they discuss, requiring readers to navigate back and forth and mentally reconstruct comment–text relationships. To preserve these relationships while maintaining reading continuity, we explored contextual comment presentation that makes comments locally accessible at the passages they address. We compared Text-End, Sentence-End, Between-Line, and Click-to-Show in a within-subject study with 24 participants using constructed comments and fixed sentence-level anchors. All three local layouts reduced initial-reading scrolling, with no significant condition effects on accuracy or elapsed task time. Sentence-End and Between-Line received higher comment-accessibility ratings than Text-End, while Click-to-Show received higher physical-demand ratings than Text-End and Sentence-End. Preferences remained distributed across all four layouts.

\end{abstract}


\wordcount{7,811 words in the main text (excluding abstract, references, and appendices); abstract 117 words}


\maketitle

\section{Introduction}
\label{sec:introduction}

Long-form content, such as articles, essays, and blog posts, has become increasingly common on online platforms. On desktop platforms such as Zhihu\footnote{\url{https://www.zhihu.com}}, Quora\footnote{\url{https://www.quora.com}}, and Medium\footnote{\url{https://medium.com}}, long-form articles often attract substantial user discussion. Readers not only read the primary text; they also turn to comments to seek alternative viewpoints, additional explanations, corrections, and reactions. Highly upvoted comments, for example, may highlight overlooked nuances or provide valuable complementary insights~\cite{YaoTXAXL15}. Comments therefore constitute an important layer of social context surrounding online reading.

However, most existing platforms spatially separate comments from the passages they discuss. Comments are typically collected in a dedicated section at the end of an article and organized by timestamp or popularity. Readers must therefore move back and forth between the source text and the comment stream to determine what a comment refers to. This separation can make comments fragmented and decontextualized, while also requiring readers to recall previously read passages when interpreting others' opinions. Moreover, comments are often redundant, noisy, and disconnected from the specific claims or passages they address. Presenting them as a single undifferentiated stream makes it difficult for readers to identify relevant perspectives and relate them to specific parts of the source text.

We use \emph{contextual comment presentation} to mean making comments locally accessible at the passages they refer to, through immediate display or markers for on-demand disclosure. Both approaches preserve a passage-level access point. Comments introduce other readers' explanations, evaluations, and reactions into the reading experience. Readers may wish to consult these perspectives alongside a passage or first consider the primary text on its own. Contextual comment presentation therefore involves coordinating access to others' interpretations with the reading of the primary text.

Other media formats illustrate the potential value of more tightly coupling audience responses with primary content. Bullet-screen comments, for instance, spatially and temporally associate user reactions with particular moments in videos, enabling viewers to perceive others' responses while consuming the content~\cite{MaC17, ChenGYT19}. Text reading, however, poses a different design challenge. Unlike video, long-form reading follows a self-paced and predominantly linear visual flow. Displaying full comments within the text, placing previews in separate blocks, and requiring readers to open a comment each organize this reading process differently. This raises a central design question: \textit{how do alternative forms of local comment presentation affect navigation, access experience, and readers' layout preferences?}

Prior research has explored various ways of improving the readability and utility of online comments. Some approaches summarize comments by extracting representative keywords~\cite{ChenLLZLCZY22, LiuHHD20} or discussion themes~\cite{WangLLCZG16, FaizahL23}, while others apply sentiment analysis to characterize emotional polarity~\cite{Pang2008,NovaSPHDP25}. These methods help readers obtain an overview of large comment collections, but they generally treat comments as a separate information space and do not explicitly preserve their relationship with the source text. Consequently, readers may still need to infer which sentence, claim, or passage a particular comment addresses.

Prior work has established the value of local access and explored its presentation trade-offs. Fluid documents compared ways of revealing supplementary information near its source and found varied preferences~\cite{Chang2000Fluid}. TaggedComments evaluated contextual access together with comment organization in tutorials~\cite{Bunt2014}, while Mullick et al. examined paragraph-level alignment and an enhanced news-comment interface~\cite{Mullick2019}. We compare three local presentation strategies with a detached list, holding the article--comment mappings constant for each article. The comparison examines which navigation benefits the local layouts share, where accessibility and operational demand differ, and how these experiences relate to individual layout choices.

Our design is also informed by longstanding annotation practices in reading. In annotated editions of ancient Chinese classics, commentary is often positioned in direct spatial relation to the original text: short annotations may appear alongside the corresponding phrase, while longer commentary may be placed at larger structural boundaries. Such practices illustrate a general principle of contextual annotation---preserving the spatial relationship between primary content and interpretive material. Inspired by this principle, we investigate whether online comments can similarly function as contextual annotations rather than as a detached discussion stream.

We implement four presentation strategies in CommentScope: Text-End (TE), Sentence-End (SE), Between-Line (BL), and Click-to-Show (CS). The prototype provides a common setting for comparing detached comments, direct local display, expandable previews, and on-demand panels. It also includes an automated comment--text alignment module for working with existing comment collections. The layout study uses fixed anchors to keep the correspondence between comments and article passages constant across conditions.

We conducted a within-subject study with 24 participants. All three local layouts reduced initial-reading scrolling relative to TE. SE and BL received higher comment-accessibility ratings than TE, while CS received higher Physical Demand ratings than TE and SE. Accuracy and elapsed task time showed no reliable condition effects. Preferences remained distributed across all four layouts. Relating individual ratings and rankings to written responses clarified what readers valued in each local layout: comparison without opening actions under SE, nearby but visually separate comments under BL, and choosing when to reveal commentary under CS. Other responses described visible comments or required interactions as disruptive. These accounts help explain why the shared navigation benefit did not produce a consistently preferred layout.

The main contributions of this work are as follows:
\begin{itemize}
    \item A controlled comparison of three local comment layouts and a detached list under fixed sentence-level anchors. The study identifies a shared reduction in initial-reading scrolling, differences in perceived accessibility and operational demand, and limited differences in accuracy and task time.
    \item A participant-level analysis of how readers evaluate the alternative layouts. Connecting all 24 participants' ratings and rankings with their written responses shows how direct comparison, visually separate commentary, and selectable access were valued or criticized. This analysis explains the different experiences accompanying local access and informs the choice of presentation beyond navigation cost alone.
\end{itemize}

\section{Related Work}

\subsection{Annotation and Assisted Reading}

Annotations connect explanations, questions, and interpretations with source material. Zhu et al.~\cite{XinranZhu} reviewed social annotation in education, identifying uses in knowledge processing, argumentation, literacy, and peer assessment. Ellis and Groth~\cite{EllisG04} developed a collaborative annotation system for data visualization. Text annotation tools also support structured labeling: WebAnno allows annotators to define linguistic annotation layers and accept or correct automatically suggested spans~\cite{Yimam2014}. These uses differ from presenting commentary for readers, but share the need to identify the material an annotation concerns.

Reading interfaces use both textual and graphical annotations. Rahman et al.~\cite{Rahman2025} organized chart annotation practices into a taxonomy and design space. Bryan et al.~\cite{Bryan2017} studied interactive text annotation generation and placement, while Kong and Agrawala~\cite{KongA12} used graphical overlays to aid chart reading and Lai et al.~\cite{Lai2020} synchronized visual annotations with textual descriptions. Other systems generate supporting representations from source content: Charagraph creates charts from data-rich paragraphs~\cite{Masson2023}, GistVis generates word-scale visualizations within documents~\cite{Zou2025}, and AutoMA generates textual annotations for time-series charts~\cite{Jiang2025}.

Assisted reading systems similarly bring supplementary information into the reading interface. Delgado et al.~\cite{DelgadoMC10} explored automated illustration for news reading, and ARSCC provides Chinese second-language learners with layered linguistic analyses~\cite{HanYMZZ20}. Paper Plain offers definitions, plain-language summaries, and question-based guidance for medical research papers~\cite{august2023paper}; other work examines generated summaries and questions for comprehension~\cite{ContrerasArguello24}. Scim allows readers to adjust the density of salient-sentence highlights during skimming~\cite{fok2022scim}, while Selenite uses generated overviews for online sensemaking~\cite{liu2024selenite}.

External material can also be linked to the source text. Papeos connects research papers with talk videos~\cite{kim2023papeos}, NewsSense links news claims with supporting or contradictory information in other articles~\cite{milbauer2023newssense}, and CiteRead integrates localized citation contexts into paper reading~\cite{rachatasumrit2022citeread}. Factful augments news with government-budget data and fact-checking support~\cite{kim2015factful}. In its laboratory study, participants produced more critical discussions and more fact-based supporting statements. Together, these systems illustrate ways to consult related information while retaining a connection to the source. Our study concerns user comments, which bring other readers' evaluations, questions, and explanations into article reading.

\subsection{Organizing and Linking Online Comments}

Comment analysis helps readers explore large collections of responses. Research has examined overall polarity~\cite{Pang2008, MaRHCTA22}, temporal changes in sentiment~\cite{Rohrdantz2012, Mao23}, common topics in news comments~\cite{ShiLBX16}, and sequential summaries of Twitter discussions~\cite{GaoLCZY14}. Visualization provides other ways to navigate these collections. Kucher et al.~\cite{Kucher2016} linked overviews of sentiment and certainty with original social-media texts, and their review surveyed visual encodings for sentiment analysis~\cite{Kucher2018}. Herder et al.~\cite{Herder2020} organized posts and comments in a graph-based interface for reflection, while Opinion Space projected comments onto a two-dimensional map to support exploration of different viewpoints~\cite{Faridani2010}.

Another line of work links a comment to the content it addresses. Hou et al.~\cite{HouLLTG17} extracted topics from news articles and comments and used a weighted classifier to align comments with news sentences. Mullick et al.~\cite{Mullick2019} developed paragraph-level alignment and an interface for reading and contributing comments at the corresponding paragraphs. Beyond text, Discursive Patinas associates comments with reader-selected chart regions and displays traces of earlier selections~\cite{Kauer2025}. These approaches establish contextual connections that readers can use to interpret commentary.

Organizing comments, assigning textual anchors, and presenting linked comments address related but distinct tasks. Our layout study focuses on presentation: a given article's comments and anchors remain fixed while their display changes. The prototype's separate alignment evaluation addresses how comments are assigned to passages; the user study compares how readers access comments once those connections have been established.

\subsection{Local Presentation of Annotations and Comments}

Fluid documents explored how supplementary information could be revealed near its source through changes to document layout. Chang et al.~\cite{Chang2000Fluid} described an observational comparison of interline, overlay, and margin presentations with footnotes and pop-up windows. Glosses near their anchors remained open for shorter intervals, and subjective preferences varied across techniques. This work already identified both the value of proximity and the different experiences associated with local presentation. Our study extends this line of comparison to article comments, examining navigation and task outcomes together with perceived accessibility, operational demand, and preference.

TaggedComments~\cite{Bunt2014} enabled users to categorize tutorial comments and pin them to relevant locations. Its evaluation compared this enhanced interface with a conventional comment interface during tutorial completion. Participants found comments more useful and helpful comments easier to locate, and preferred the enhanced interface, although some noted added visual complexity. Task performance did not differ significantly. These results demonstrate that improved comment access and subjective experience need not be accompanied by a task-performance gain. The evaluation concerned a package of comment-organization and access features; our comparison examines differences among alternative local layouts as well as their shared benefits relative to a detached list.

Mullick et al.~\cite{Mullick2019} evaluated automatic comment alignment and a paragraph-level news-comment interface against the original news website. Their work connects textual association with local access and comment contribution. In our study, the mappings for each article are held constant across four presentation conditions. This provides a common basis for comparing complete layouts without differences in automatic alignment quality.

We compare full comments after sentences (SE), expandable previews between text segments (BL), and comments revealed in a selectable panel (CS), using a detached end-of-article list (TE) as the reference condition. We examine which navigation benefits are shared across the local layouts, where accessibility and operational demand differ, and how readers evaluate these experiences when choosing a layout. Relating each participant's ratings and ranking to their written responses allows us to examine favorable and unfavorable accounts of direct display, separate comment blocks, and selectable access. The comparison brings these forms of local presentation into the same reading tasks, rather than evaluating local access through a single enhanced interface.

\section{Design of Contextual Comment Presentation}

Here, we describe three established contextual comment presentation strategies that vary in spatial proximity, visibility, and interaction requirements, together with a conventional Text-End baseline. We implement these strategies in CommentScope, a research prototype that supports contextual access to comments at sentence, paragraph, and article levels.

\subsection{Design Goals}

Our layout comparison is guided by two goals derived from the challenges described in Section~\ref{sec:introduction}.

\textbf{G1: Preserve comment--text context.}
Comments should remain closely connected to the passages they discuss, reducing the need for readers to mentally reconstruct their relationship with the source text.

\textbf{G2: Balance comment visibility and reading continuity.}
Comments should be sufficiently visible to support discovery and interpretation, while avoiding excessive interruption to the primary reading flow.

\subsection{Contextual Comment Presentation Strategies}
\label{sec:presentation_strategies}

\begin{figure}[ht]
  \centering
  \includegraphics[width=\linewidth]{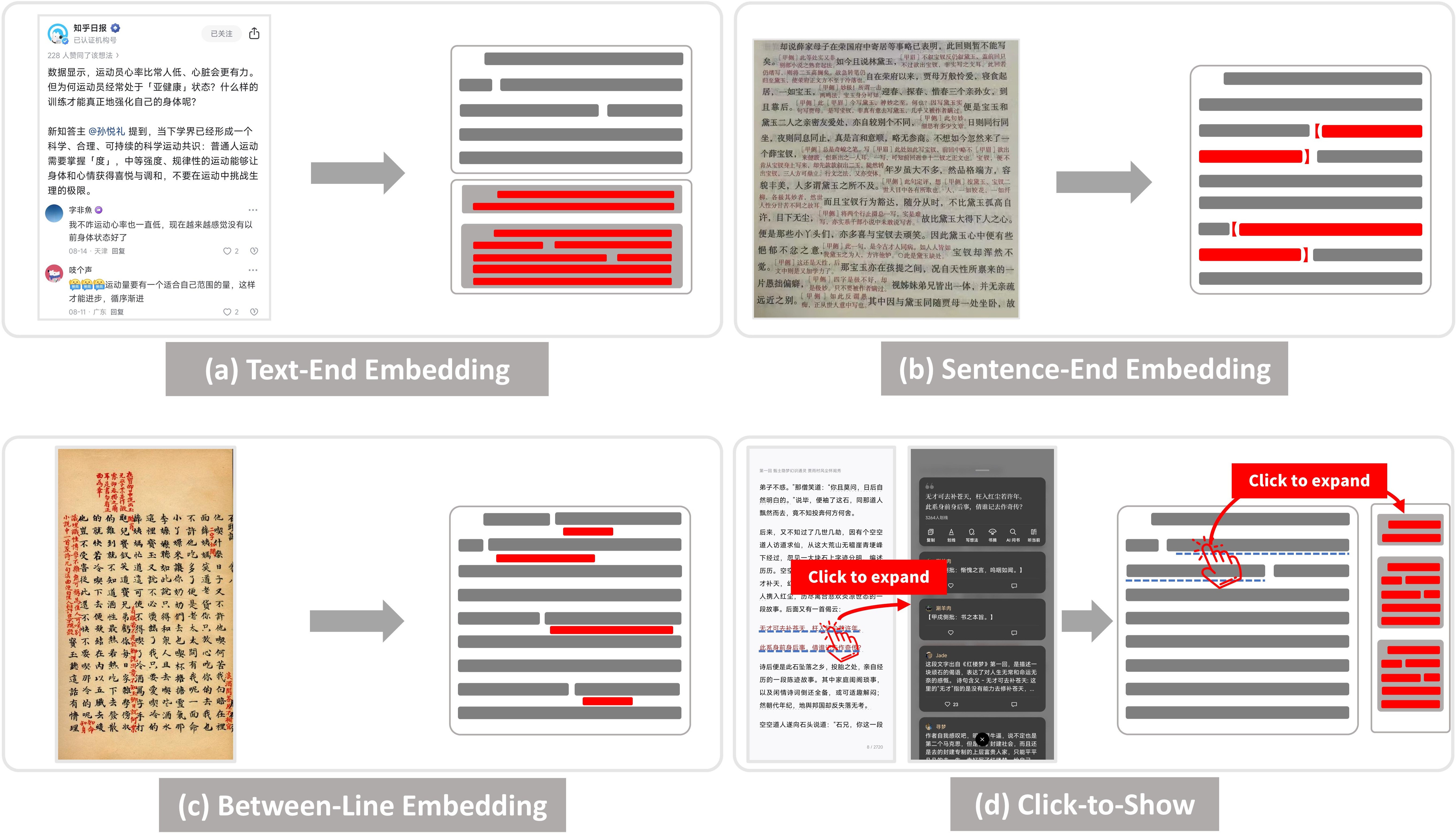}
  \caption{Design references and schematics for four comment presentation strategies. In each pair, the example on the left motivates the schematic on the right; gray strokes represent article text and red strokes represent comments. (a) Text-End collects comments in a separate section after the article. (b) Sentence-End places comments immediately after their anchor sentences. (c) Between-Line inserts comments between text segments. (d) Click-to-Show reveals a separate comment panel when a marked passage is selected. The first strategy separates text and commentary; the other three preserve a local access point, with different visibility and interaction requirements.}
  \Description{A schematic comparison of four comment layouts: comments collected after the article, a comment placed after its anchor sentence, a comment block inserted between text segments, and comments revealed by selecting a marked sentence.}
  \label{fig:schematic_diagram}
\end{figure}

We derived the strategies from existing comment and annotation patterns. Text-End follows the conventional detached comment list (Fig.~\ref{fig:schematic_diagram}a); Sentence-End places an annotation after the corresponding sentence (Fig.~\ref{fig:schematic_diagram}b); Between-Line inserts commentary between text segments (Fig.~\ref{fig:schematic_diagram}c); and Click-to-Show reveals comments through an anchor selected by the reader (Fig.~\ref{fig:schematic_diagram}d). Sentence-End (SE), Between-Line (BL), and Click-to-Show (CS) share local anchoring: each provides access to comments from their referenced passages. SE displays comment text directly, BL displays a comment or an expandable preview, and CS displays a marker for on-demand access. Text-End (TE) serves as the detached baseline.

\begin{itemize}

    \item \textbf{Text-End Baseline (TE; Fig.~\ref{fig:schematic_diagram}a).}
    Following the conventional layout used by many online platforms, comments are presented in a separate section after the article. This design keeps comments outside the article body, allowing readers to read the text before reaching the comment section. It also spatially separates comments from the passages they discuss and may require readers to navigate back and forth.

    \item \textbf{Sentence-End (SE, Fig.~\ref{fig:schematic_diagram}b).}
    The most-liked comment associated with a sentence is displayed immediately after that sentence. This strategy maximizes spatial proximity while keeping commentary within the local reading flow. Compared with larger embedded blocks, the lightweight inline presentation is intended to make comments immediately accessible without substantially altering the structure of the article.

    \item \textbf{Between-Line (BL, Fig.~\ref{fig:schematic_diagram}c).}
    The most-liked comment associated with a sentence is presented as a visually distinct block between adjacent text segments. Longer comments can be expanded on click. The stronger visual separation makes commentary highly visible and difficult to overlook, but occupies more space within the article and may introduce greater interruption to continuous reading.

    \item \textbf{Click-to-Show (CS, Fig.~\ref{fig:schematic_diagram}d).}
    Sentence-level comment text remains hidden by default and is revealed only when readers explicitly select a marked sentence. This strategy keeps comment text out of the article layout, but introduces additional interaction and requires readers to actively decide when to inspect commentary.

\end{itemize}

Table~\ref{tab:presentation_attributes} summarizes the access and disclosure attributes of the four layouts. Sentence-End emphasizes immediate access and spatial proximity, Between-Line presents comments as separate blocks, and Click-to-Show provides selective access through local markers. Each layout combines several attributes; for example, CS combines hidden comment text with presentation in a side panel. The study compares these complete layouts using a shared source-material set with predefined passage references. Section~\ref{sec:study_materials} describes the study materials.

\begin{table}[htbp]
  \centering
  \caption{Access and disclosure attributes of the four presentation conditions. SE and BL display comment text directly; BL also supports expansion of longer comments.}
  \label{tab:presentation_attributes}
  \small
  \begin{tabular}{p{0.08\linewidth}p{0.19\linewidth}p{0.27\linewidth}p{0.31\linewidth}}
    \toprule
    Layout & Passage-level entry point & Default comment presentation & Access action \\
    \midrule
    TE & No & Cards after the article & Scroll to the list; revisit passages as needed \\
    CS & Yes & Hidden; marked passages & Select a marker to reveal comments in a side panel \\
    SE & Yes & Comment after its anchor sentence & Read the displayed comment \\
    BL & Yes & Separate block or preview between text segments & Read the preview; expand longer comments \\
    \bottomrule
  \end{tabular}
\end{table}

\subsection{Research Prototype}

We implement the layouts in CommentScope, a web-based research prototype with a central article area and contextual comment access. The controlled study assigns one layout per article and uses fixed sentence-level anchors. Its interface records reading, navigation, comment interactions, answers, and ratings.

\begin{figure}[htbp]
  \centering
  \includegraphics[width=\linewidth]{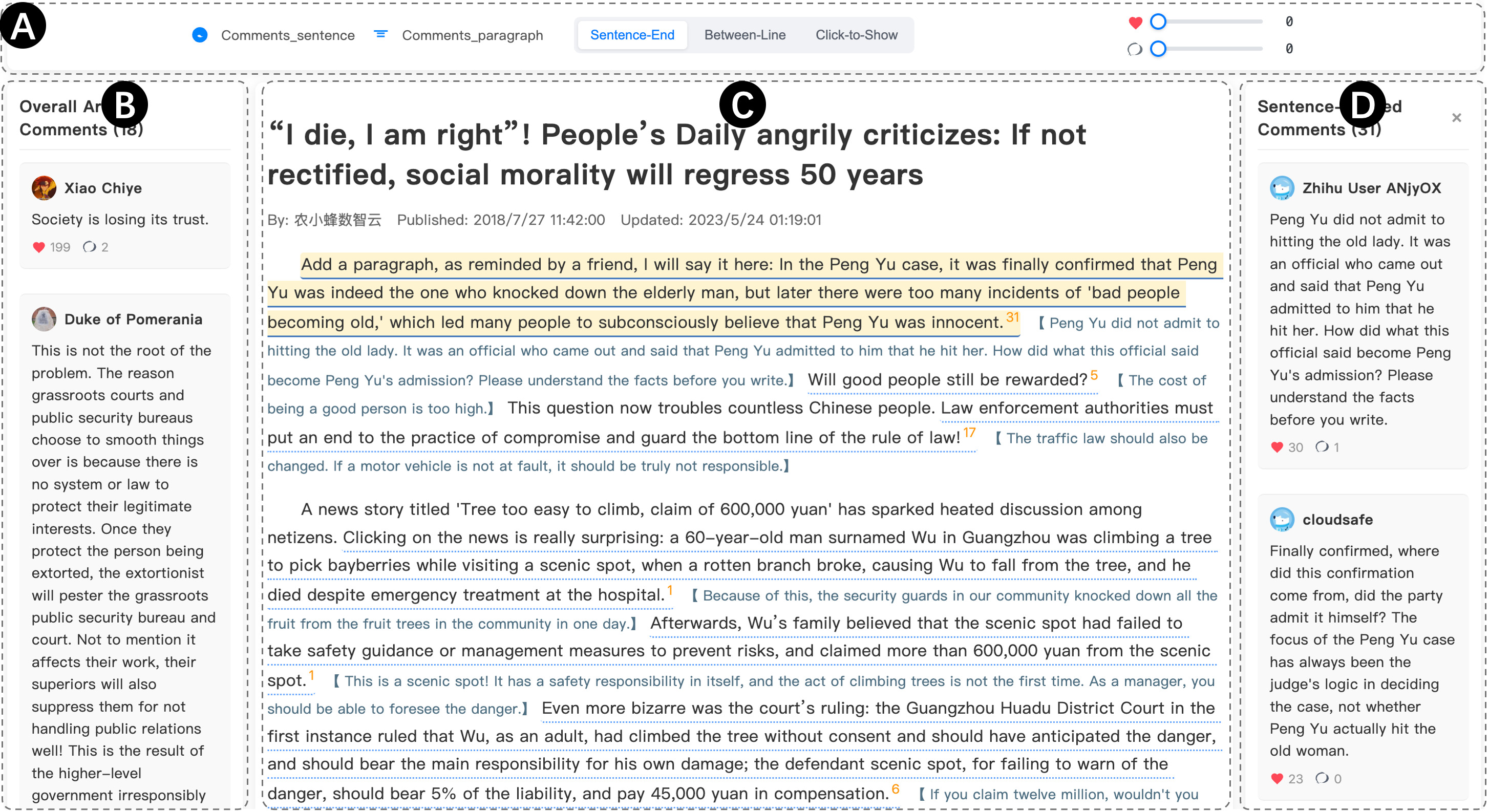}
  \caption{The broader CommentScope prototype. (A) Layout, annotation-level, and filtering controls. (B) Article-level comments. (C) Central article with commented sentences and the selected anchor. (D) Comments associated with that anchor. The controlled study used assigned sentence-level layouts with fixed anchors.}
  \Description{A labeled screenshot shows the top control bar, left global-comment panel, central article with highlighted anchor text, and right sentence-comment panel.}
  \label{fig:system-overview}
\end{figure}

The broader prototype also supports sentence-, paragraph-, and article-level comments, layout switching, and filters based on likes and replies. Figure~\ref{fig:system-overview} shows this interface, and Appendix~\ref{app:prototype_details} describes the additional controls. These features support exploration of existing comment collections; the layout comparison uses assigned configurations without switching or filtering.

\section{Comment--Text Alignment in the Prototype}
\label{sec:evaluation}

Contextual presentation requires comments to be associated with the article passages they address. CommentScope includes an alignment module that predicts a sentence-level, paragraph-level, or global reference and returns the corresponding anchor in structured JSON. The layout study uses predefined anchors; the following technical evaluation examines automatic alignment for existing comment collections.

We annotated 1,000 comments from ten Zhihu articles and split the data by article: eight articles with 800 comments for training and two articles with 200 comments for testing. Two authors independently labeled reference levels and target segments. Agreement on reference levels was assessed on 20\% of the dataset, yielding Cohen's $\kappa=0.91$. We fine-tuned Qwen3-8B with LoRA and compared it with the zero-shot base model using the same alignment instructions. Appendix~\ref{app:alignment_details} provides the annotation and training details, followed by the alignment prompt.

Position Exact Match (EM) requires both the reference level and the textual anchor to match the human annotation. On the 200 held-out comments, EM increased from 0.52 for the zero-shot model to 0.82 after fine-tuning (Table~\ref{tab:evaluation_results}). This result supports the use of task-specific training for anchor prediction on the held-out articles. How remaining alignment errors affect reading is a question for an end-to-end evaluation.

\begin{table}[htbp]
  \centering
  \caption{Comment--text alignment on 200 held-out comments from two articles. Position EM requires an exact match on both the reference level and textual anchor.}
  \label{tab:evaluation_results}
  \begin{tabular}{lc}
    \toprule
    \textbf{Method} & \textbf{Position EM} \\
    \midrule
    Base Model (Zero-shot) & 0.52 \\
    Fine-tuned Model & 0.82 \\
    \bottomrule
  \end{tabular}
\end{table}

\section{Evaluation of Comment Layouts}

We conducted a within-subject study of four presentation conditions using constructed comments with predefined passage references. Three layouts provided local comment access through immediate display or selectable markers; Text-End used a detached list. We examined task performance, initial-reading scrolling, perceived workload, and preference.

\begin{figure}[htbp]
  \centering
  \includegraphics[width=\linewidth]{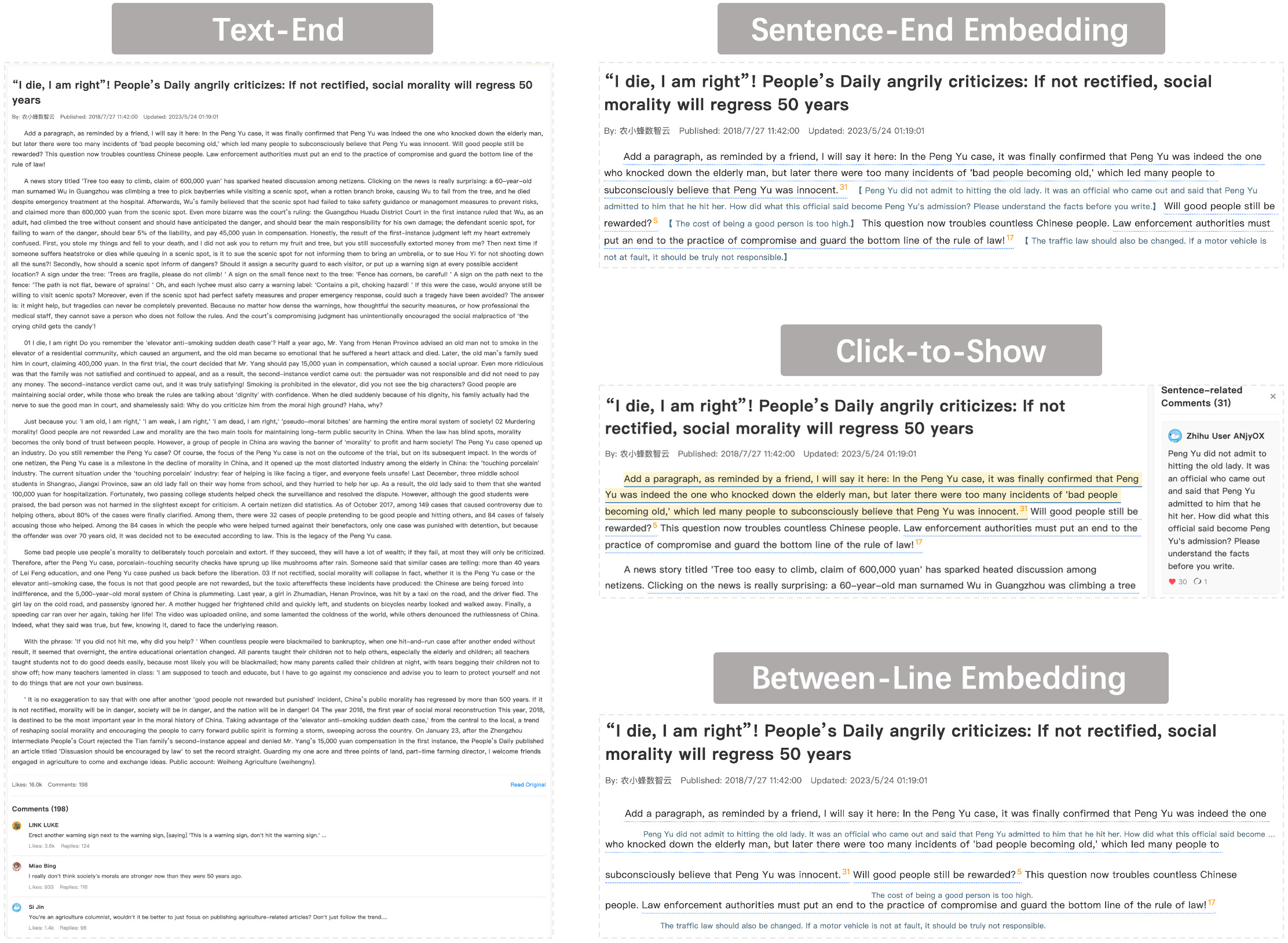}
  \caption{Four prototype layouts illustrated with the same example article. Left: Text-End (TE) collects comments below the complete article. Upper right: Sentence-End (SE) places a comment immediately after its anchor sentence. Middle right: Click-to-Show (CS) highlights a selected passage and reveals its comments in a side panel. Lower right: Between-Line (BL) inserts comments between text segments. These views illustrate presentation and interaction differences; the example article is not one of the controlled study materials.}
  \Description{One tall view on the left shows a detached comment list below an article. Three views on the right show sentence-end comments at the top, an expanded click-to-show panel in the middle, and between-line comments at the bottom.}
  \label{fig:embedding_methods}
\end{figure}

\subsection{User Study}

We organize the evaluation around four research questions:

\begin{itemize}
    \item \textbf{RQ1:} How does comment presentation affect initial-reading navigation and elapsed task time?
    \item \textbf{RQ2:} How does comment presentation affect accuracy and recorded response time in article comprehension, comment recognition, and comment--text integration?
    \item \textbf{RQ3:} How do readers evaluate comment accessibility, reading continuity, and workload across layouts?
    \item \textbf{RQ4:} Which layouts do readers prefer, and what aspects of comment access and reading experience do they value in their explanations?
\end{itemize}

Each participant experienced all four conditions, with one article assigned to each condition. We used an eight-sequence Latin-square design with three participants per sequence to balance article order, interface order, and article--interface combinations. Each article--condition combination and each condition--order position included six participants (Appendix~\ref{app:study_sequences}). The four conditions, illustrated in Fig.~\ref{fig:embedding_methods}, were:

\begin{itemize}
    \item \textbf{Text-End (TE):} All comments were collected as cards after the article. The article itself contained no contextual comment presentation.
    \item \textbf{Click-to-Show (CS):} Markers appeared beside sentences associated with comments. Comment text remained hidden until participants selected a marker, after which it appeared in a side panel.
    \item \textbf{Sentence-End (SE):} The associated comment was displayed directly after its anchor sentence in a visually distinct inline style.
    \item \textbf{Between-Line (BL):} The associated comment appeared as a separate block between adjacent text segments, with an initial single-line preview that could be clicked to expand.
\end{itemize}

\subsection{Materials}
\label{sec:study_materials}

We used four Chinese long-form articles from saved Zhihu records, each paired with ten comments prepared through model-assisted drafting and revision. Following researcher review, the comments were stored and loaded as fixed study materials. Appendix~\ref{app:study_materials} details the article sources, comment preparation, material characteristics, and presentation rules.

Each article was paired with six single-choice questions, with one correct answer per item:
\begin{itemize}
    \item \textbf{Comment recognition (two items):} Participants judged whether a stated viewpoint had appeared in the comments by selecting one of two options, ``seen'' or ``not seen.'' Each article included one target viewpoint from its prepared comment set and one distractor absent from that set.
    \item \textbf{Article comprehension (two items):} Participants selected one of four answers to a question about the article's claims or explanations, testing understanding of the primary text.
    \item \textbf{Comment--text integration (two items):} Participants selected one of four answers about the relationship between a comment and the article content it addressed, testing whether they could connect commentary to its textual context.
\end{itemize}
For a given article, the text, comments, comment--text anchors, questions, and correct answers were held constant across presentation conditions. Only the comment layout and its associated access mechanism changed.

\subsection{Participants}

The study included 24 undergraduate students from diverse academic backgrounds, including humanities, computing, engineering, and business. Each participant completed all four conditions.

\subsection{Apparatus}

We implemented the study as a web-based experiment on desktop computers. The study used assigned layouts and predefined anchor quotations, without the prototype's layout-switching or filtering controls. Participants read the article, answered questions, and completed subjective evaluations within the same interface. The system automatically logged reading time, scroll behavior, comment interactions, question responses, and response time.

\subsection{Procedure and Tasks}

\subsubsection{Article Reading and Question Tasks}

For each article, an introduction page identified the assigned interface and displayed its schematic. Participants then read at their own pace, without a forced time limit. After reading, they answered the six questions described in Section~\ref{sec:study_materials}, presented one at a time. The article and comments were hidden during recognition questions, which participants answered from memory. During comprehension questions and questions about comment--text relationships, participants could close the question panel, consult the article and comments, and reopen the panel to answer. The system recorded reading activity, navigation, comment interactions, selected answers, and timing information throughout these stages.

\subsubsection{Subjective Evaluation and Preference}

After answering each article's questions, participants completed eight seven-point ratings: six items based on the NASA Task Load Index (NASA-TLX) dimensions and two subjective evaluations of reading continuity and comment accessibility. The item definitions and scale directions are specified in Section~\ref{sec:study_measures}.

After completing all four articles, participants ranked the four presentation strategies from most to least preferred and explained their first choice. They then answered four open-ended questions about preferred layouts, reading interruption, comment--text search strategies, and desired interface improvements.

\subsection{Measures and Analysis}
\label{sec:study_measures}

The measures were defined as follows:
\begin{itemize}
    \item \textbf{Comment Recognition Accuracy (CRA):} The proportion of correct responses to the two recognition questions, measuring whether participants recognized viewpoints expressed in the comments.
    \item \textbf{Article Comprehension Accuracy (ACA):} The proportion of correct responses to the two comprehension questions, measuring understanding of the article content.
    \item \textbf{Comment--Text Integration Accuracy (CTIA):} The proportion of correct responses to the two questions about how a comment related to the article content it addressed.
    \item \textbf{Comment--Text Integration Response Time (CTIRT):} The average answering time for those two questions. For each question, timing began when the question panel was last opened and ended when the answer was submitted.
    \item \textbf{Reading Continuity (RC):} A single-item rating from 1 to 7 assessing whether the interface supported continuous reading; higher scores indicated a more favorable experience.
    \item \textbf{Comment Accessibility (CA):} A single-item rating from 1 to 7 assessing whether relevant comments were easy to find and read; higher scores indicated a more favorable experience.
\end{itemize}

\textbf{Reading and response time.} Initial reading time covers article reading before the questions. Recorded response time runs from the final opening of a question panel to answer submission, excluding earlier viewing and article consultation. We averaged the two recorded times for each question type; the average for the comment--text relationship questions is CTIRT. We summed all six recorded times to obtain \emph{recorded answer time}.

\emph{Elapsed task time} spans the start of initial reading to the final answer submission, including consultation, transitions, and pauses but excluding ratings and feedback.

\textbf{Navigation and comment access.} During initial article reading, we recorded scroll-event counts and normalized scroll distance, together with comment-panel opening and closing actions in CS. Normalized scroll distance was the total distance of all upward and downward scrolling divided by the page height recorded at the end of reading; a value of 1.0 corresponds to one page-height of cumulative scrolling.

\textbf{Workload and preference.} The workload questionnaire covered the six NASA-TLX dimensions~\cite{HartStaveland1988TLX}: Mental Demand, Physical Demand, Temporal Demand, Performance, Effort, and Frustration. Each item used a seven-point scale (1 = very low, 7 = very high). Physical Demand covered clicking and scrolling. The Performance item asked how difficult participants found performing the task; higher scores indicated greater difficulty. Participants ranked the four layouts from 1 (most preferred) to 4 (least preferred). We also counted how often their first-choice layout had their highest RC or CA, lowest Physical Demand, fewest scroll events, or shortest elapsed task time, allowing ties.

\textbf{Statistical analysis.} We used tie-corrected Friedman tests for within-participant condition comparisons. Significant omnibus results were followed by two-sided paired Wilcoxon signed-rank tests, with Holm correction across the six condition pairs for each outcome and $\alpha=.05$. Stage-specific timing, article-specific summaries, and cross-metric comparisons are exploratory. In the text, $M$ denotes the arithmetic mean and $Mdn$ the median. Figure captions specify the summary statistic and confidence interval used in each plot; condition-wise intervals describe uncertainty within each condition.

\textbf{Exploratory analysis of ratings and written responses.} Following the condition-level analyses, we compared each participant's first-choice layout with their ratings across all four conditions. We recorded whether that layout received their highest CA, highest RC, or lowest Physical Demand, allowing ties, and summarized these matches for each first-choice group. We then examined the written explanations alongside the same participant's ratings, considering both supporters and critics of each layout. Appendix~\ref{app:participant_ratings} also includes a supplementary CS--SE comparison of Physical Demand and RC, calculated as the CS rating minus the SE rating for each participant.

We examined the 24 first-choice explanations and 96 open-ended answers, giving 120 entries from 24 participants. We retained each entry's source and read the five entries together for each participant. We grouped stated reasons concerning access effort, visible commentary, reading sequence, and visual separation, and retained familiarity and perceived task utility as additional explanations. Reported consultation strategies were examined separately. Table~\ref{tab:response_groups} specifies the statements included in each group. An entry could express several concerns. We compared the accounts with the same participant's rankings and ratings, including differing evaluations across questions. Quotations are translated from Chinese.

\begin{table}[htbp]
\centering
\small
\caption{Grouping rules for the exploratory written-response analysis. The groups describe overlapping concerns expressed in the responses.}
\label{tab:response_groups}
\begin{tabular}{@{}p{0.24\linewidth}p{0.48\linewidth}p{0.22\linewidth}@{}}
\toprule
\textbf{Concern} & \textbf{Included statements} & \textbf{Examples} \\
\midrule
Access effort & Scrolling, searching, opening, or switching described as helpful or disruptive. & P01, P05, P14 \\
\addlinespace
Visible commentary & Comments described as drawing attention or interrupting primary-text reading. & P11, P16, P22 \\
\addlinespace
Reading sequence & Completing the article first, interleaving commentary, or choosing whether and when to consult it. & P03, P04, P05 \\
\addlinespace
Visual separation & Distinguishing source from commentary through blocks, spacing, typography, or boundaries. & P07, P21, P23 \\
\addlinespace
Familiarity and task utility & Habit or perceived help with understanding and answering. & P04, P08, P18 \\
\addlinespace
Consultation strategies & Reported keyword lookup, memory, revisiting, selective comparison, or foregoing a search. & P01, P03, P11, P21 \\
\bottomrule
\end{tabular}
\end{table}

Brief answers naming only a layout were recorded as choices, and answers such as ``observation'' or ``clicking'' were recorded as unspecified strategies. Interface suggestions were recorded as requested changes.

Final rankings determined first-choice counts. Four participants (P02, P09, P12, and P20) selected a different layout in the separate open preference answer. Rankings and open preference answers were recorded separately.

\subsection{Results}

\subsubsection{Utility: Navigation and Task Time}

\begin{figure}[htbp]
  \centering
  \includegraphics[width=\linewidth]{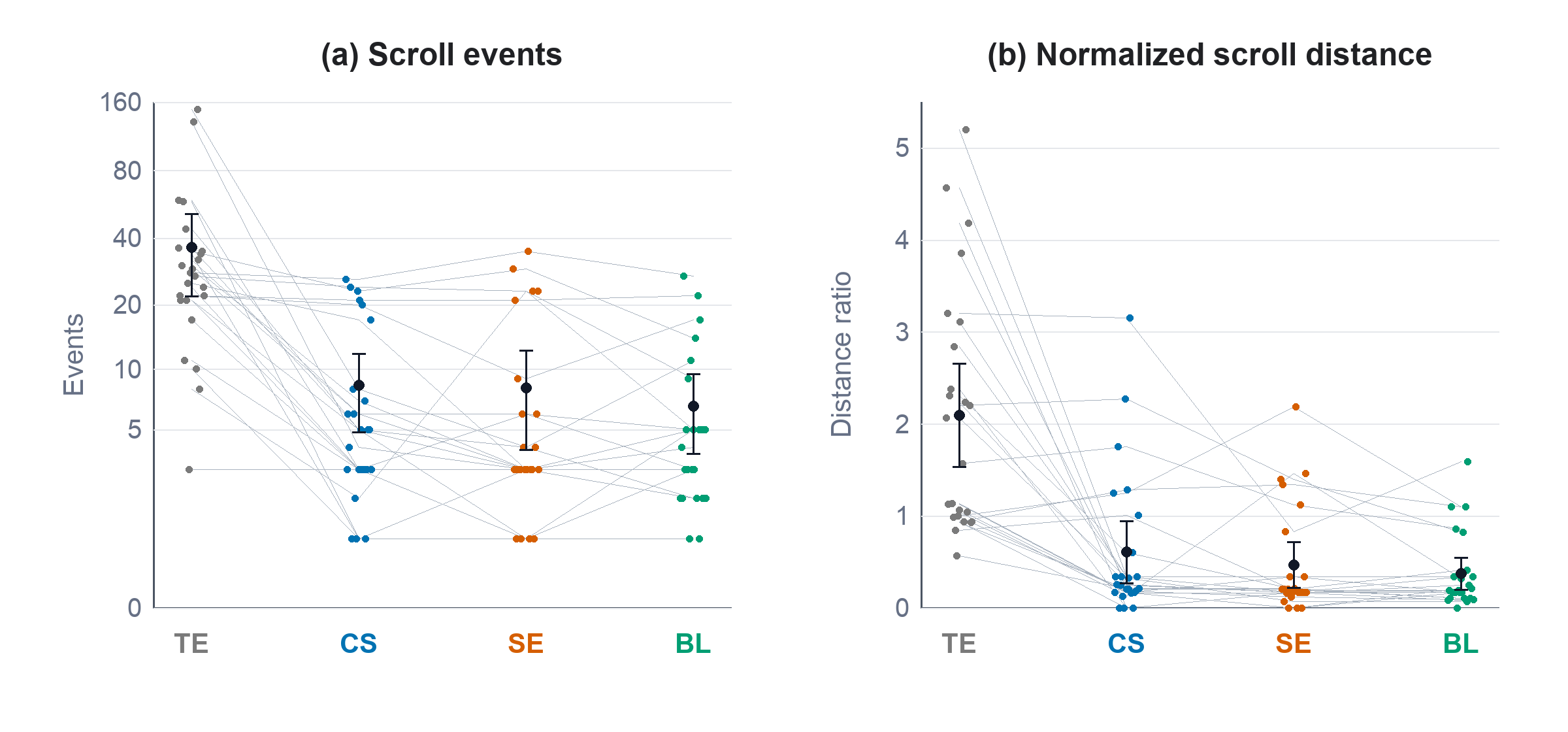}
  \caption{Navigation during initial article reading, with 24 participants per condition. 
  (a) Number of logged scroll events. (b) Total distance of all upward and downward scrolling divided by the page height recorded at the end of reading; 1.0 corresponds to one page-height of cumulative scrolling. Colored points are participant observations, horizontally offset for visibility; pale lines connect the same participant across conditions. Black circles and whiskers show means and condition-wise 95\% $t$ confidence intervals. Panel (a) uses a $\log(1+x)$ vertical scale with ticks in original event counts; panel (b) is linear. 
  TE: Text-End; CS: Click-to-Show; SE: Sentence-End; BL: Between-Line.}
  \Description{Panel a shows scroll-event counts, and panel b shows normalized scrolling distance during initial reading. TE has higher values than the three contextual layouts. Each panel shows 24 participant observations per condition, paired lines, and mean confidence intervals.}
  \label{fig:navigation_cost}
\end{figure}

We analyzed initial-reading scrolling and elapsed task time to examine how the four layouts supported navigation and task completion. Participants produced more scroll events under TE ($M=36.46$) than under CS ($M=8.33$), SE ($M=8.08$), and BL ($M=6.58$; Fig.~\ref{fig:navigation_cost}a), $\chi^2(3)=39.12$, $p<.001$. Normalized scroll distance showed the same pattern, with mean values of 2.09 for TE, 0.61 for CS, 0.47 for SE, and 0.37 for BL (Fig.~\ref{fig:navigation_cost}b), $\chi^2(3)=28.51$, $p<.001$. Holm-adjusted comparisons were significant between TE and each locally anchored layout, but not among CS, SE, and BL. At the participant level, 23 of 24 participants produced fewer scroll events under CS than TE, 21 under SE, and 23 under BL. TE also had the highest article-level mean for both navigation measures in all four articles, although the ordering among locally anchored layouts varied (Appendix~\ref{app:article_navigation}). In CS, SE, and BL, readers could access comments from the passages they addressed. This local access may explain the lower scrolling counts, as readers could consult commentary without moving to a separate list below the article.

\begin{figure}[htbp]
  \centering
  \includegraphics[width=\linewidth]{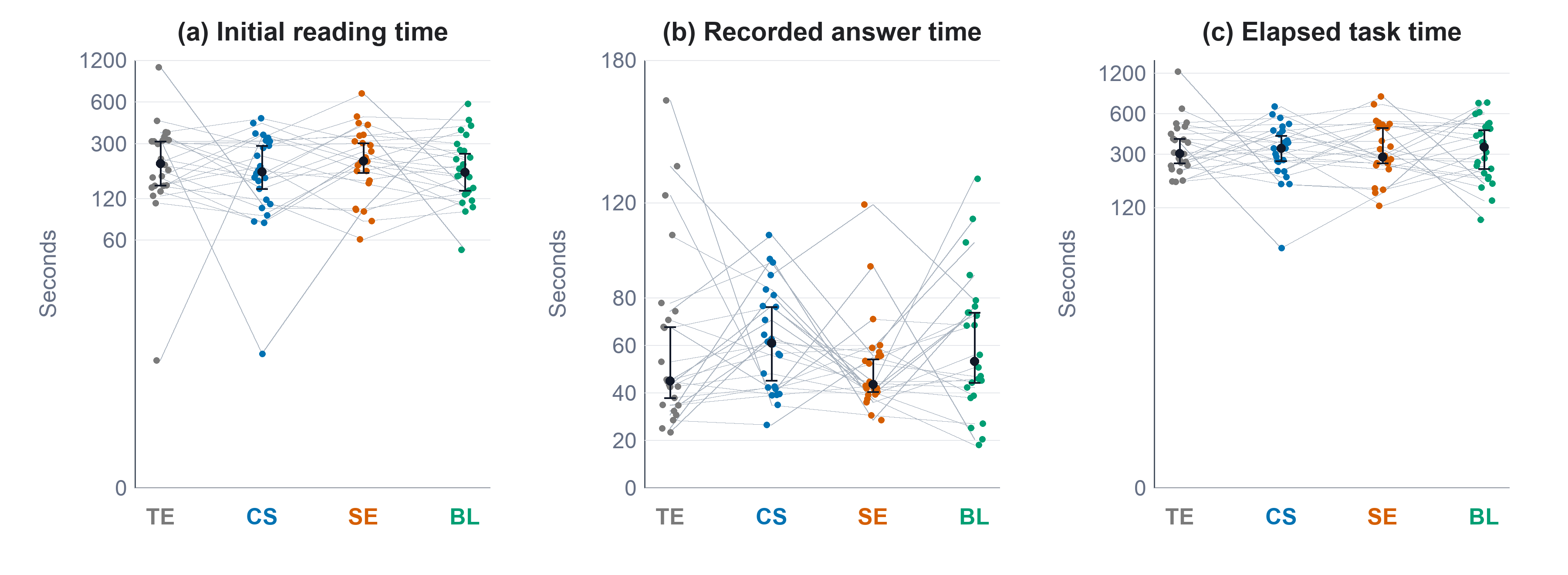}
  \caption{Time measures across the reading-and-question task. (a) Initial reading time. (b) Sum of the six recorded answering intervals, each measured from the final question-panel opening to submission. (c) Complete elapsed time from initial reading start to final question submission, excluding ratings. Colored points show 24 participant observations per condition; pale lines connect participants. Black circles are medians, with condition-wise percentile-bootstrap 95\% confidence intervals from 2,000 participant resamples. Panels (a) and (c) use $\log(1+x)$ scales with ticks in seconds; panel (b) is linear. TE: Text-End; CS: Click-to-Show; SE: Sentence-End; BL: Between-Line.}
  \Description{Three panels show initial reading time, recorded answer time, and elapsed task time. Each panel contains paired participant observations, condition medians, and bootstrap confidence intervals. Reading and elapsed task times use logarithmic scales; recorded answer time uses a linear scale.}
  \label{fig:task_time_decomposition}
\end{figure}

We next examined whether the reduction in scrolling was accompanied by faster task completion. Initial reading time and recorded answer time showed no significant condition effects (Fig.~\ref{fig:task_time_decomposition}a--b; $p=.648$ and $p=.278$), and elapsed task time also showed no significant condition effect (Fig.~\ref{fig:task_time_decomposition}c; $\chi^2(3)=1.25$, $p=.741$). Mean elapsed times were 374.16~s for TE, 348.47~s for CS, 355.81~s for SE, and 363.23~s for BL. CS had the shortest mean, whereas SE had the shortest median (285.91~s); only 9 of 24 participants completed the task faster under SE than TE, compared with 21 of 24 who produced fewer scroll events under SE than TE. Together, these results show a consistent reduction in initial-reading scrolling under the local layouts, while the timing measures showed no reliable improvement in completion speed.

\subsubsection{Utility: Accuracy and Response Time}

We examined answer accuracy to evaluate how well each layout supported comment recognition, article comprehension, and comment--text integration. Accuracy remained high across conditions. CRA ranged from 79.2\% under CS to 89.6\% under BL, with SE at 85.4\% and TE at 83.3\%, but the condition effect was not significant, $\chi^2(3)=1.67$, $p=.644$. ACA ranged from 93.8\% to 100.0\% ($\chi^2(3)=5.18$, $p=.159$), and CTIA from 95.8\% to 100.0\% ($\chi^2(3)=4.00$, $p=.261$). Participants answered most comprehension and integration questions correctly with consultation available, leaving little room for the layouts to differ on these measures. Recognition from memory showed greater variation, although no layout had a significant advantage.

\begin{figure}[htbp]
  \centering
  \includegraphics[width=\linewidth]{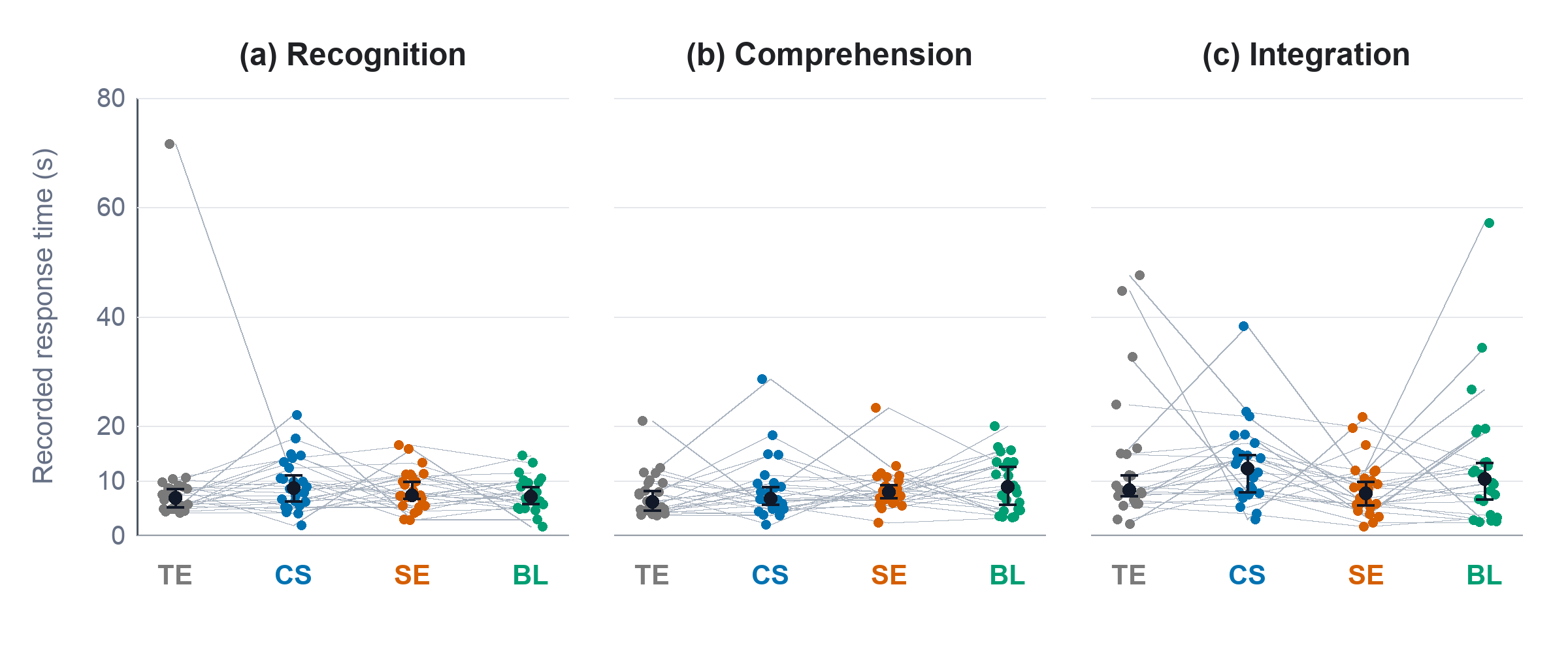}
  \caption{Recorded response time by question type: (a) comment-recognition response time, (b) article-comprehension response time, and (c) Comment--Text Integration Response Time (CTIRT). Each colored point is a participant's mean over two recorded item intervals measured from final panel opening to answer submission. Each condition contains 24 points, with pale lines connecting the same participant within each panel. Black circles are medians of the participant-level means, with percentile-bootstrap 95\% confidence intervals from 2,000 participant resamples. The three equal-sized panels share a linear time scale labeled at left. Only (c) had an omnibus effect; no CTIRT pairwise contrast survived Holm correction. TE: Text-End; CS: Click-to-Show; SE: Sentence-End; BL: Between-Line.}
  \Description{Three equal-sized panels in one horizontal row show mean per-item recorded response times for comment recognition, article comprehension, and comment--text integration. A shared vertical axis at left ranges from zero to eighty seconds. Each panel contains individual paired observations, condition medians, and bootstrap confidence intervals.}
  \label{fig:task_type_response_time}
\end{figure}

We also compared the recorded answering intervals for the three question types. Recognition time and comprehension time did not differ significantly (Fig.~\ref{fig:task_type_response_time}a--b; $p=.475$ and $p=.183$). CTIRT showed an omnibus effect (Fig.~\ref{fig:task_type_response_time}c; $\chi^2(3)=9.15$, $p=.027$), with the lowest mean under SE (8.58~s), compared with 12.98~s under CS, 13.12~s under BL, and 13.28~s under TE. However, no pairwise contrast survived Holm correction; the smallest adjusted $p$-value was .091 for CS versus SE. These intervals cover the final question-panel opening to submission; earlier question viewing and article consultation are included only in elapsed task time. Thus, SE had a shorter mean final answering interval for integration questions, while accuracy and adjusted pairwise timing comparisons showed no reliable advantage for any layout.

\subsubsection{Perception: Comment Accessibility and Workload}

\begin{figure}[htbp]
  \centering
  \includegraphics[width=\linewidth]{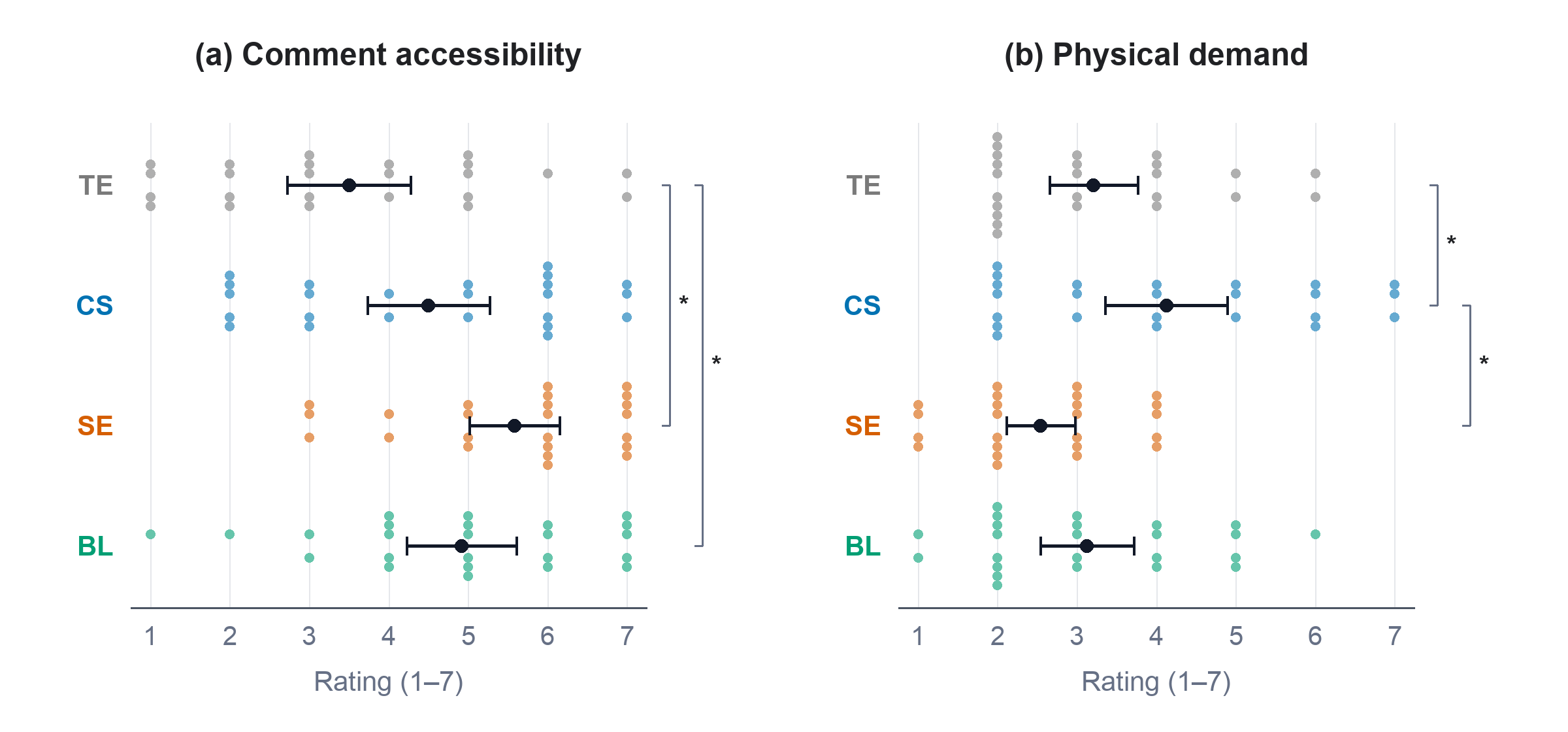}
  \caption{Two distinct aspects of subjective experience. (a) Comment Accessibility (CA), from 1 (strongly disagree) to 7 (strongly agree) that relevant comments are easy to find and read; higher is more favorable. (b) Physical Demand, from 1 (very low) to 7 (very high) perceived operational demand; lower is more favorable. Each condition occupies a row with 24 individual ratings. Horizontal position encodes rating; colored points at the same rating are separated vertically for visibility. Black circles and horizontal whiskers show means and condition-wise 95\% $t$ confidence intervals. Brackets to the right of each panel mark Holm-adjusted paired comparisons with $p<.05$: TE--SE and TE--BL in (a), TE--CS and CS--SE in (b). TE: Text-End; CS: Click-to-Show; SE: Sentence-End; BL: Between-Line.}
  \Description{Two horizontal rating plots, with conditions ordered TE, CS, SE, and BL from top to bottom. Panel a shows comment-accessibility ratings with significant TE--SE and TE--BL comparisons. Panel b shows physical-demand ratings with significant TE--CS and CS--SE comparisons. Colored dots show all 24 ratings per condition, separated vertically at identical ratings without connecting lines. Black mean points have horizontal confidence intervals; significant paired comparisons are bracketed outside the rating scale.}
  \label{fig:subjective_tradeoff}
\end{figure}

To examine the experience of finding and reading comments, we analyzed Comment Accessibility (CA), Reading Continuity (RC), and the six workload ratings. CA differed significantly across conditions (Fig.~\ref{fig:subjective_tradeoff}a), $\chi^2(3)=15.99$, $p=.001$. Its means were 5.58 for SE, 4.92 for BL, 4.50 for CS, and 3.50 for TE. Holm-adjusted comparisons showed higher CA for SE than TE ($p=.005$) and for BL than TE ($p=.034$), indicating that participants found comments easier to locate and read in these two layouts.

Physical Demand showed a different pattern from scrolling. Despite its low scroll count, CS received the highest Physical Demand rating (4.13), exceeding TE (3.21; adjusted $p=.037$) and SE (2.54; adjusted $p=.015$; Fig.~\ref{fig:subjective_tradeoff}b). During initial reading, participants made 6.46 comment-opening actions and 1.96 closing actions on average under CS. Opening comments added an interaction step in CS, whereas SE displayed comments directly and BL provided an inline preview. This difference offers a possible explanation for the higher operational demand reported under CS despite its low scrolling counts. The remaining workload dimensions and RC showed no reliable condition effects (all $p\geq.205$). Overall, SE and BL made comments easier to access than TE, while CS combined nearby access with the additional effort of revealing comment content.

\subsubsection{Preference and Reported Reading Experience}

\begin{figure}[htbp]
  \centering
  \includegraphics[width=\linewidth]{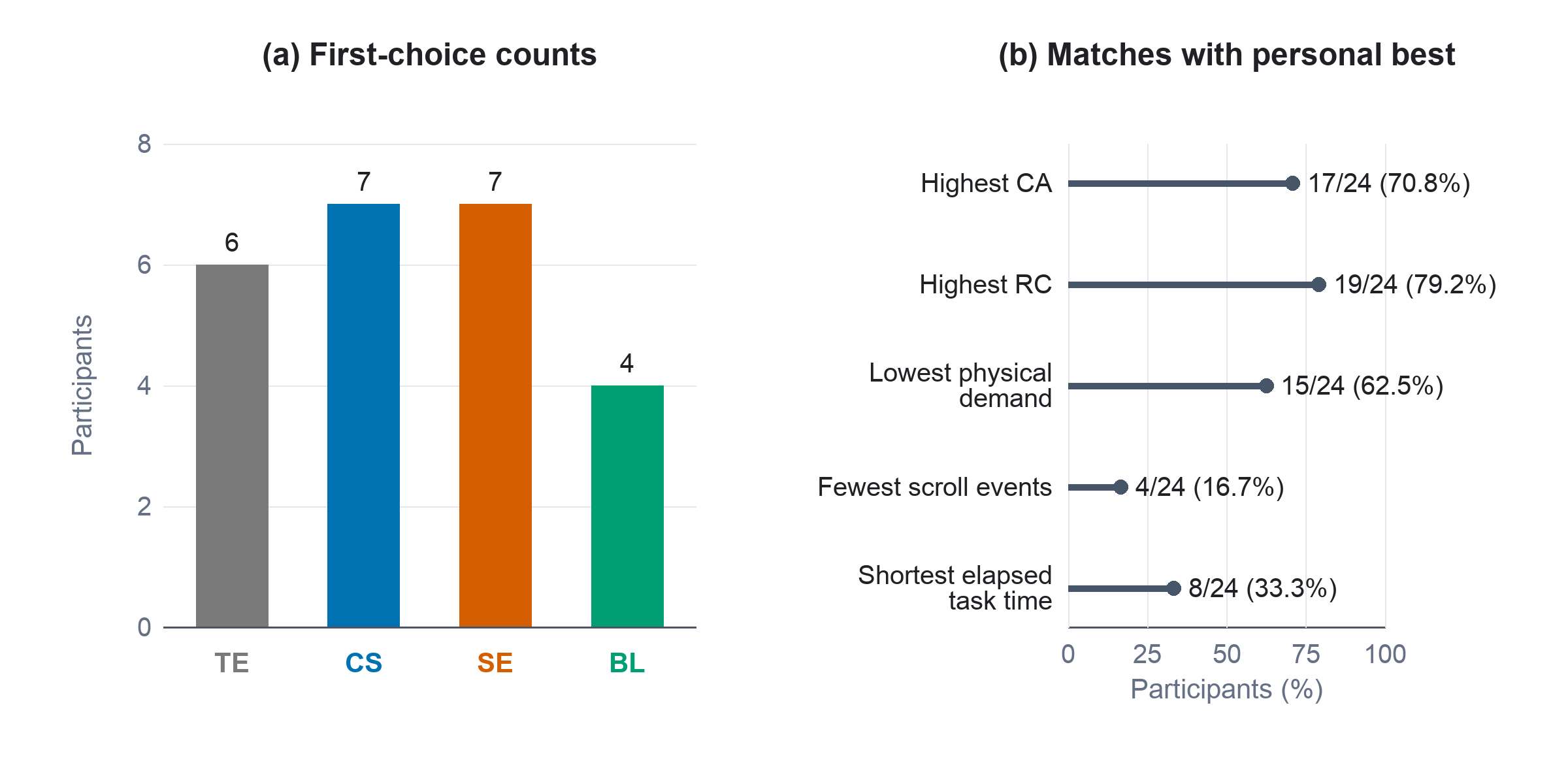}
  \caption{Preference and within-participant alignment. (a) Colored vertical bars show first-choice counts out of 24 participants for Text-End (TE), Click-to-Show (CS), Sentence-End (SE), and Between-Line (BL); labels above the bars give counts. Each participant contributes one first choice. (b) The percentage whose first choice also attained their highest Comment Accessibility (CA), highest Reading Continuity (RC), lowest Physical Demand, fewest initial-reading scroll events, or shortest elapsed task time, measured from initial reading start to final question submission. Labels in (b) give counts and percentages out of 24; matches include tied best conditions. These percentages describe matches with individual best scores, including ties; differences in tie frequency limit comparisons across measures.}
  \Description{Panel a uses zero-based vertical bars to show six first choices for TE, seven each for CS and SE, and four for BL. Panel b retains a horizontal percentage plot showing 17 accessibility matches, 19 continuity matches, 15 physical-demand matches, four scroll-count matches, and eight complete-task-time matches, allowing tied best conditions.}
  \label{fig:preference_alignment}
\end{figure}

We analyzed the final rankings alongside participants' ratings and written responses to examine what they valued in each layout. SE and CS each received seven first-choice selections, followed by TE with six and BL with four (Fig.~\ref{fig:preference_alignment}a). Mean preference ranks were 2.25 for SE, 2.50 for CS and BL, and 2.75 for TE; the overall difference was not significant, $\chi^2(3)=1.80$, $p=.615$. Thus, the shared navigation benefit of local presentation did not produce a consistently preferred layout.

A participant's preferred layout also received their highest RC rating in 19 of 24 cases, highest CA in 17, and lowest Physical Demand in 15 (Fig.~\ref{fig:preference_alignment}b). The corresponding counts were four for the fewest scroll events and eight for the shortest elapsed task time. Table~\ref{tab:preference_rating_matches} breaks down the subjective-rating matches by first choice, using the same comparison across all four layouts for every participant. Matches include tied best values; their frequency differs across measures. We used the written responses to examine what participants valued in these choices.

\begin{table}[htbp]
\centering
\small
\caption{Participants whose first-choice layout also received their best rating among the four layouts. Each participant belongs to one row and can contribute to several rating columns. Counts include ties; $n$ is the number selecting that layout first.}
\label{tab:preference_rating_matches}
\begin{tabular}{lrrrr}
\toprule
\textbf{First choice} & \textbf{n} & \textbf{Highest CA} & \textbf{Highest RC} & \textbf{Lowest Physical Demand} \\
\midrule
SE & 7 & 7 & 6 & 6 \\
BL & 4 & 3 & 4 & 2 \\
CS & 7 & 4 & 5 & 2 \\
TE & 6 & 3 & 4 & 5 \\
\bottomrule
\end{tabular}
\end{table}

\textbf{SE: Direct comparison without an opening step.} All seven SE supporters gave it their highest CA rating, and six gave it their lowest Physical Demand and highest RC ratings, allowing ties. Their explanations emphasized access to comments at the relevant sentence. P01 valued having ``no extra operations,'' describing required interactions as interrupting thought and the distance to the CS panel as making comparison difficult. P05 similarly emphasized the direct correspondence and fewer scrolling and clicking actions. These accounts connect SE's appeal to considering a passage and its commentary together without first opening another view.

This convenience could matter even when the participant gave SE a less favorable rating elsewhere. P09 ranked SE first and rated its CA above CS (6 versus 2), but rated its RC below CS (2 versus 6). The ranking explanation emphasized matching comments to sentences and the convenience of considering them together. Conversely, P11, who preferred CS, reported that visible SE comments attracted attention even when unwanted. Direct display therefore made comparison convenient for some readers while bringing commentary into the reading flow before others wanted to consult it.

\textbf{BL: Nearby comments in separate blocks.} All four BL supporters gave it their highest RC rating; three also gave it their highest CA and two their lowest Physical Demand, allowing ties. P06 valued considering comments with the article without much scrolling or clicking. P20 emphasized distinguishing the article from the comments, ranking BL first even though its CA rating was lower than SE's (4 versus 6). These explanations connect BL's appeal to keeping comments nearby while separating them from the primary text.

P21 described BL's independent blocks as allowing comments to be read alongside the article, but found that SE encouraged continuing into the comment rather than the next part of the source. This distinction appeared in the written account despite identical SE and BL ratings for CA (6), RC (6), and Physical Demand (2). P23 likewise valued convenient lookup under BL, while describing SE as interrupting the article and repeated CS clicks as ``annoying and tedious.'' The block layout was not universally welcomed: P07, who preferred TE, found BL's paragraph indentation confusing. Thus, BL's spatial separation was part of how participants evaluated nearby commentary, not simply whether comments were visible.

\textbf{CS: Choosing when to reveal commentary.} Five of the seven CS supporters gave it their highest RC rating, four their highest CA, and two their lowest Physical Demand, allowing ties. P13, P18, and P22 rated CS higher than SE in both Physical Demand and RC, and described a clear reading layout, fewer inline interruptions, or access at passages of interest. P18 also cited familiarity. These accounts suggest why additional actions were acceptable to some readers: comments appeared when requested rather than being displayed within the article from the outset.

Control over viewing did not always accompany a higher continuity score. P03 valued the freedom to open and close comments, saying, ``I can open them when I want to read them and close them when I do not,'' but rated RC lower under CS than SE (4 versus 7). Other participants disliked the required actions: P10 described the extra action before seeing a comment as distracting, and P17 reported difficulty reading with CS and a reluctance to open comments. Both ranked TE first. Selectable access could therefore be valued as a choice over comment viewing or experienced as an extra step in reaching the content.

\textbf{TE: Reading the article before the comments.} TE remained the first choice for six participants. Four gave it their highest RC rating, three their highest CA, and five their lowest Physical Demand, allowing ties. P04 and P10 wanted to finish the article before considering comments, and P04 also cited familiarity. These accounts explain the appeal of a detached list despite its greater scrolling: it kept commentary out of the article reading sequence.

Reported consultation strategies also varied within these preferences. Participants described looking up keywords (P03, P04, and P24), relying on memory (P07, P09, P19, and P21), or revisiting the source for a comment of interest (P01). P05 preferred SE for direct correspondence but reported reading the article first and consulting comments during a second pass. Layout preference therefore did not prescribe one consultation sequence.

Together, the ratings and written responses explain why the three local layouts did not lead to one preferred experience. SE supporters emphasized comparison without extra opening actions. BL supporters valued nearby comments that remained visually distinct from the article. CS supporters emphasized choosing when to reveal commentary, while other readers found the required actions disruptive. These evaluations show how the manner of presenting a nearby comment mattered alongside the ease of reaching it. Appendix~\ref{app:participant_ratings} provides all participants' ratings and an additional CS--SE comparison.

\FloatBarrier

\section{Discussion and Future Work}

\subsection{Shared Navigation Benefits of Local Presentation}

The clearest common benefit of SE, BL, and CS was reduced scrolling during initial reading. Each provided access to comments from the passages they addressed, rather than requiring readers to move to a detached list. This pattern appeared despite differences in whether comments were shown in full, previewed, or opened in a panel. The comparison supports local comment access across these configurations, rather than identifying one local layout as the best way to reduce navigation.

The navigation benefit was not accompanied by reliable gains in accuracy or elapsed task time. A reading session includes considering the article, interpreting comments, consulting the source, and answering questions; reducing scrolling changes only part of that process. The results therefore favor evaluating navigation and task performance separately. Scroll counts describe movement through the page, while complete task intervals show whether readers finish sooner. The demonstrated benefit concerns navigation, not a general improvement in comprehension or reading speed.

\subsection{Different Experiences of Nearby Comments}

Local disclosure and contextual comment access have already been explored through supplementary glosses, tutorial comments, and paragraph-linked news comments~\cite{Chang2000Fluid,Bunt2014,Mullick2019}. Our comparison brings three local presentation strategies into the same reading tasks and connects their navigation and access outcomes with individual evaluations. SE and BL improved perceived accessibility relative to TE, whereas CS combined reduced scrolling with greater operational demand than TE and SE. The written accounts further distinguish how readers experienced these layouts, even when their ratings were similar.

\textbf{Sentence-end display makes commentary available during comparison.} SE removes the step of opening a comment before considering it with the source sentence. Its higher accessibility than TE and lower operational demand than CS were accompanied by accounts of convenient comparison. Other readers described the same sentence-end placement as drawing attention away from the article. For this layout, a central design concern is how to make a comment easy to consult while keeping the continuation of the source text easy to follow. Typography, spacing, and the boundary between source and comment are relevant to that concern.

\textbf{Between-line presentation gives nearby commentary a separate place.} BL provided local previews and higher accessibility than TE, while supporters emphasized independent blocks and the distinction between source and commentary. This adds a consideration beyond whether comments are immediately visible: readers also evaluate how they are separated from the article. The contrasting account of confusing paragraph indentation shows that inserting a block can itself affect the perceived structure. These observations motivate further comparisons of preview length, spacing, and expansion behavior within BL.

\textbf{Selectable panels make access depend on an opening action.} CS lets readers continue through the article without first displaying the comment text. Some valued this control despite greater perceived effort, while others found the repeated actions disruptive. Its operational cost and its selective access are both relevant to the experience. Design work on this layout should consider the effort of opening comments and comparing the panel with the source, as well as the ability to leave commentary hidden.

Together, these findings place the three layouts on a common basis: all provide nearby access, but they differ in how commentary enters the reading process. Direct display, separate previews, and selectable panels should therefore be evaluated through both access measures and readers' accounts of using them. The assigned layouts in this study combine placement, visibility, and interaction; future comparisons could vary these features separately to determine which changes improve each presentation strategy.

\subsection{Extending the Comparison to Denser Discussion}

The study used ten comments per article. In denser discussions, multiple long comments at one passage would add more text to SE, more previews or expansion actions to BL, and more material to browse in CS or TE. Varying comment density and clustering would therefore directly test when the navigation benefit of local access is outweighed by the space or actions needed to present comments.

Comment selection becomes important under these conditions. The broader prototype can surface the most-liked comment inline, but popularity alone may leave less common perspectives out of view. Replies also introduce relationships among comments, beyond each comment's connection to the article. Future evaluations could compare comment-selection and expansion rules while measuring both the effort of reaching relevant discussion and the viewpoints readers encounter. These extensions would test the layouts under more complex discussion conditions rather than simply adding similar article topics.

\subsection{Limitations and Future Directions}

Our study involved 24 undergraduate students and four Chinese long-form commentary articles. The texts varied in topic and length, and each participant read a different article in each condition. Counterbalancing distributed article and order combinations across layouts, but participant-level contrasts also include variation between articles. Larger studies with explicit modeling of article and order effects would help establish how these patterns extend to other readers and materials. The comments were model-assisted constructions with fixed references to the text; their preparation is described in Appendix~\ref{app:study_materials}.

Each question type contained two items, and ACA and CTIA were near ceiling. These measures had limited sensitivity to layout differences. Recognition was memory-based, whereas comprehension and integration allowed consultation. Recorded response times covered only the final question-panel opening to submission; elapsed task time covered the full reading-and-question interval. The absence of reliable accuracy and elapsed-time effects does not establish equivalent performance across layouts.

Holm correction was applied within each outcome. The analyses of timing by stage, matches between preferences and ratings, and individual rating differences are exploratory. The alignment counts include tied best values, with different tie frequencies across measures. RC was a single-item rating, and the explanations came from brief retrospective written answers. These materials describe what participants valued, rather than continuously observing when each comment was read. Longer interviews and time-resolved observation would help examine how access preferences relate to actual consultation behavior.

The user study used fixed anchors and assigned layouts, so its findings concern presentation rather than automatic alignment, filtering, or layout switching. The separate alignment evaluation concerns Chinese textual data. Testing the complete system would require examining how alignment errors, comment selection, and reader-controlled disclosure affect the experience under realistic use.

\section{Conclusion}

We compared Text-End, Sentence-End, Between-Line, and Click-to-Show in a within-subject study with 24 undergraduate participants using constructed comments and fixed sentence-level anchors. All three local layouts reduced initial-reading scrolling. SE and BL received higher comment-accessibility ratings than TE, while CS received higher Physical Demand ratings than TE and SE. Accuracy and elapsed task time showed no reliable condition effects, and preferences remained distributed across all four layouts.

Relating participants' ratings and rankings to their written responses clarified the appeal and difficulties of each local presentation. SE offered comparison without opening actions, but sentence-end comments could draw attention away from continued article reading. BL accounts emphasized the visual distinction between nearby commentary and the source, with differing views of the block arrangement. CS offered control over when comments appeared, but the required actions were also experienced as disruptive. These findings show that a shared navigation benefit can accompany different access experiences. Designing local comment interfaces requires attention to how commentary is distinguished from the source and made available for consultation, not only how close it is to the relevant passage.

\section*{Acknowledgments}

We sincerely thank all participants who took part in our user study for their
time, effort, and valuable feedback.

\section*{Funding}

The authors report that no specific funding was obtained for this work.
\section*{Ethics Approval}
This study was conducted with the approval of the Experimental Ethics Review
Committee of Jiangsu Ocean University (no
approval number was assigned).

\section*{Disclosure Statement}

The authors report there are no competing interests to declare.

\section*{Data Availability Statement}

The data supporting the findings of this study are available from the
corresponding author upon reasonable request. Raw participant-level data are
not publicly available because of participant privacy and ethical
considerations. Source materials derived from third-party Zhihu records are
subject to applicable source and platform restrictions.

\section*{Declaration of Generative AI Use}

Generative AI tools were used as research tools for comment--text alignment and
model-assisted drafting of the fixed study comments. The alignment experiments
used Qwen3-8B with LoRA and the LLaMA-Factory framework. All model outputs used
in the study and manuscript were reviewed and verified by the authors. All
substantive content, analyses, interpretations, and conclusions were developed
and verified by the authors, who take full responsibility for the integrity of
the manuscript and its references. No generative AI tool is an author.

\bibliographystyle{ACM-Reference-Format}
\bibliography{sample-base}

\appendix
\section{Supplementary Details}
\label{sec:appendix_prompts}
This appendix provides full participant ratings, study sequences, article-specific navigation summaries, and material details. It also describes the broader prototype and alignment module. The main text reports the four-layout preference analysis and written accounts.

\subsection{Participant Ratings and Preference}
\label{app:participant_ratings}

Table~\ref{tab:participant_ratings} reports Comment Accessibility (CA), Reading Continuity (RC), and Physical Demand for all 24 participants in each layout. First choices come from the final ranking. Table~\ref{tab:preference_rating_matches} in the main text summarizes matches between these choices and individual best ratings, allowing ties.

\begin{table}[htbp]
  \centering
  \small
  \setlength{\tabcolsep}{3.5pt}
  \caption{All 24 participants' first-choice layouts and seven-point ratings for Comment Accessibility, Reading Continuity, and Physical Demand. Higher accessibility and continuity scores are more favorable; lower demand scores indicate less operational effort. First choices come from the final ranking.}
  \label{tab:participant_ratings}
  \begin{tabular*}{\linewidth}{@{\extracolsep{\fill}}ll*{12}{r}@{}}
    \toprule
    & & \multicolumn{4}{c}{Comment Accessibility} & \multicolumn{4}{c}{Reading Continuity} & \multicolumn{4}{c}{Physical Demand} \\
    \cmidrule(lr){3-6} \cmidrule(lr){7-10} \cmidrule(lr){11-14}
    Participant & First choice & TE & CS & SE & BL & TE & CS & SE & BL & TE & CS & SE & BL \\
    \midrule
    P01 & SE & 3 & 7 & 7 & 7 & 4 & 7 & 7 & 6 & 2 & 2 & 1 & 2 \\
    P02 & CS & 4 & 3 & 5 & 2 & 4 & 3 & 4 & 2 & 4 & 4 & 4 & 5 \\
    P03 & CS & 5 & 6 & 6 & 5 & 6 & 4 & 7 & 6 & 4 & 3 & 3 & 5 \\
    P04 & TE & 5 & 4 & 6 & 5 & 5 & 5 & 5 & 5 & 2 & 2 & 2 & 2 \\
    P05 & SE & 3 & 2 & 6 & 6 & 3 & 2 & 6 & 4 & 4 & 6 & 3 & 3 \\
    P06 & BL & 4 & 6 & 7 & 7 & 5 & 6 & 7 & 7 & 2 & 2 & 2 & 1 \\
    P07 & TE & 5 & 6 & 3 & 4 & 6 & 6 & 4 & 3 & 2 & 2 & 3 & 2 \\
    P08 & SE & 2 & 2 & 7 & 5 & 3 & 4 & 7 & 6 & 6 & 6 & 3 & 3 \\
    P09 & SE & 3 & 2 & 6 & 6 & 4 & 6 & 2 & 3 & 2 & 2 & 3 & 4 \\
    P10 & TE & 5 & 3 & 5 & 5 & 6 & 2 & 6 & 3 & 3 & 6 & 4 & 3 \\
    P11 & CS & 1 & 7 & 3 & 7 & 6 & 7 & 3 & 4 & 3 & 3 & 3 & 2 \\
    P12 & SE & 1 & 2 & 6 & 3 & 3 & 4 & 5 & 3 & 2 & 5 & 1 & 5 \\
    P13 & CS & 2 & 5 & 5 & 4 & 4 & 6 & 5 & 3 & 3 & 3 & 2 & 4 \\
    P14 & SE & 2 & 3 & 7 & 7 & 2 & 3 & 7 & 6 & 5 & 5 & 1 & 4 \\
    P15 & TE & 7 & 3 & 7 & 4 & 1 & 4 & 3 & 3 & 4 & 4 & 3 & 4 \\
    P16 & CS & 4 & 5 & 7 & 7 & 6 & 6 & 6 & 6 & 4 & 4 & 4 & 3 \\
    P17 & TE & 7 & 5 & 5 & 5 & 6 & 3 & 4 & 3 & 3 & 6 & 4 & 5 \\
    P18 & CS & 1 & 6 & 7 & 4 & 1 & 5 & 3 & 2 & 5 & 7 & 2 & 1 \\
    P19 & TE & 3 & 6 & 4 & 3 & 4 & 7 & 2 & 4 & 2 & 2 & 4 & 2 \\
    P20 & BL & 1 & 2 & 6 & 4 & 4 & 1 & 2 & 5 & 2 & 4 & 1 & 2 \\
    P21 & BL & 6 & 6 & 6 & 6 & 5 & 5 & 6 & 6 & 2 & 2 & 2 & 2 \\
    P22 & CS & 2 & 7 & 3 & 1 & 2 & 7 & 1 & 1 & 3 & 5 & 2 & 2 \\
    P23 & BL & 5 & 6 & 4 & 6 & 4 & 2 & 5 & 5 & 2 & 7 & 2 & 3 \\
    P24 & SE & 3 & 4 & 6 & 5 & 3 & 4 & 5 & 5 & 6 & 7 & 2 & 6 \\
    \bottomrule
  \end{tabular*}
\end{table}

\subsubsection{Supplementary CS--SE Rating Comparison}

\begin{figure}[htbp]
  \centering
  \includegraphics[width=\linewidth]{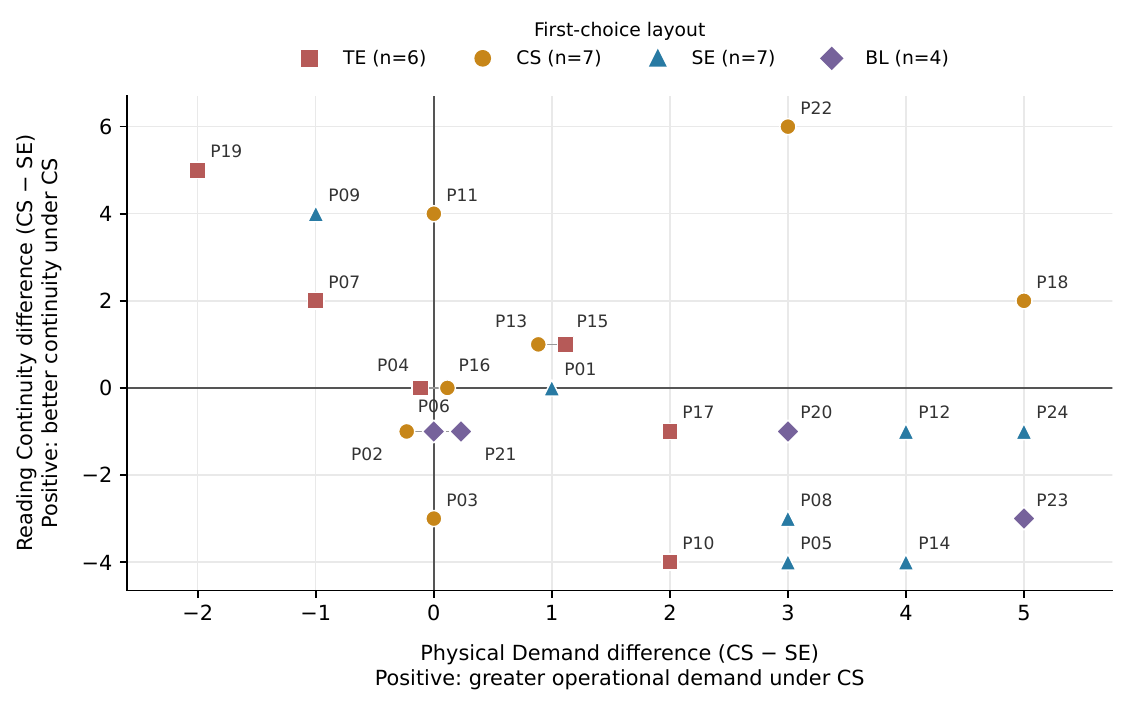}
  \caption{Within-participant differences between CS and SE for all 24 participants. The horizontal axis shows Physical Demand under CS minus SE; the vertical axis shows Reading Continuity (RC) under CS minus SE. Both original ratings use seven-point scales. Positive horizontal values indicate greater demand under CS, and positive vertical values indicate better continuity under CS. Color and shape identify the first-choice layout. Participants with identical difference pairs are separated slightly along the horizontal axis for visibility.}
  \Description{A scatterplot with participant labels shows all 24 CS-minus-SE rating pairs. Four participants have positive differences on both measures: P13, P18, and P22 preferred CS, while P15 preferred TE. The remaining CS supporters include one with higher RC, one with equal RC, and two with lower RC. The other preference groups span several combinations of demand and continuity.}
  \label{fig:demand_continuity_all}
\end{figure}

Figure~\ref{fig:demand_continuity_all} compares CS and SE for all 24 participants. Fourteen rated CS higher in Physical Demand, seven rated the layouts equally, and three rated SE higher. For RC, eight rated CS higher, three gave equal ratings, and thirteen rated SE higher. Among CS supporters, three rated CS higher in both demand and continuity; the other four rated demand equally, with one giving CS higher RC, one equal RC, and two lower RC. One TE supporter also rated CS higher than SE in both demand and continuity. The figure retains this pairwise detail alongside the four-layout preference analysis in the main text.

\subsection{Additional Prototype Features}
\label{app:prototype_details}

The broader prototype supports exploration at several textual levels. These controls complement the four layouts described in Section~\ref{sec:presentation_strategies}.

\subsubsection{Article Content Area}
\label{article_content_area}

The article content area preserves the original paragraph structure while augmenting the text with lightweight visual cues for comment exploration~\cite{BeckW17}.

At the sentence level, sentences associated with comments are marked by an orange comment count and a blue dashed underline. Selecting a sentence highlights it and reveals all comments associated with that sentence in the right-side panel. Depending on the selected presentation strategy, the most-liked comment associated with the sentence may additionally appear at the sentence end or between text segments, while sentence-level comment text in Click-to-Show is revealed only through the panel.

\begin{figure}[h]
  \centering
  \includegraphics[width=0.8\linewidth]{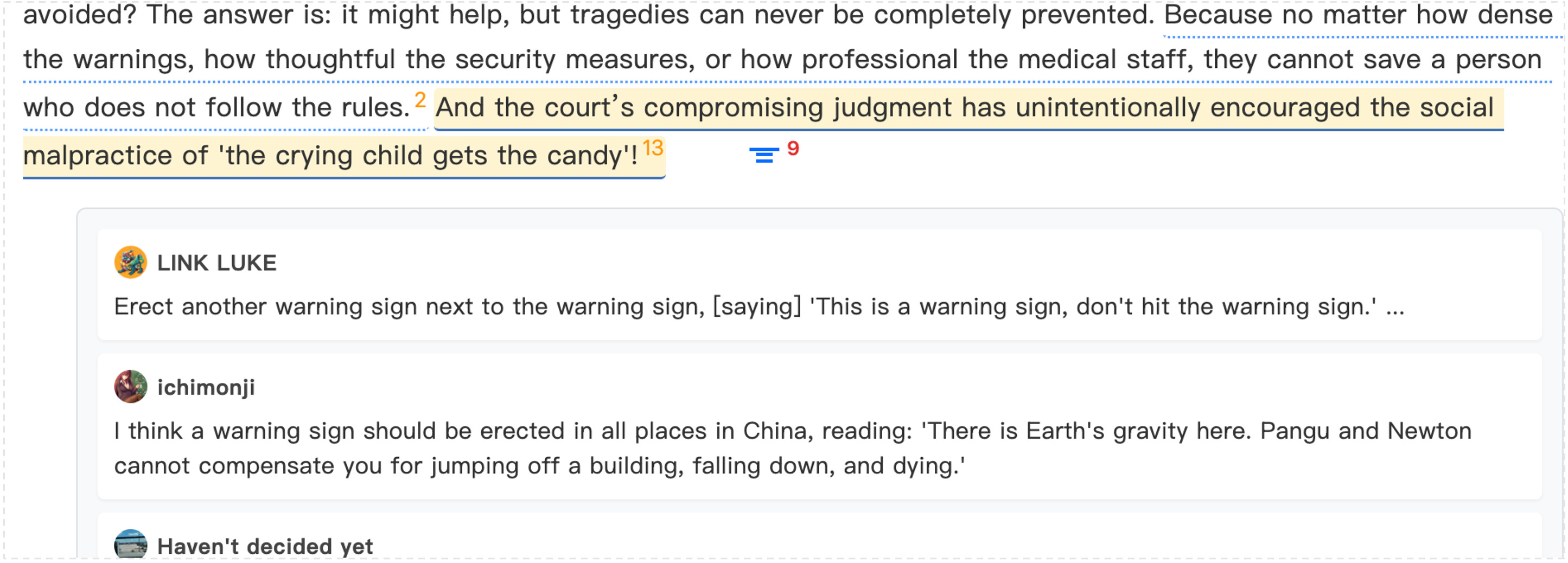}
  \caption{Expanded paragraph-level comments in the prototype. The blue paragraph marker and adjacent red count indicate nine comments associated with the paragraph; selecting the marker reveals the comment cards below the text. Blue underlines and orange counts separately identify sentence-level comments. Paragraph-level disclosure provides access to broader commentary without requiring every comment to be placed after a single sentence.}
  \Description{An expanded paragraph-comment area below article text. A blue paragraph icon has a red count of nine; comment cards appear underneath, while orange numbers mark separate sentence-comment counts.}
  \label{fig:paragraph-level}
\end{figure}

At the paragraph level, CommentScope provides collapsible access to comments associated with an entire paragraph (Fig.~\ref{fig:paragraph-level}). A numeric indicator communicates comment volume, allowing readers to identify sections that attract concentrated discussion.

\subsubsection{Multi-Level Comment Panels}

CommentScope provides two side panels for accessing commentary at different levels of granularity.

The fixed \textbf{Global-level Comment Panel} presents comments that refer to the article as a whole. Comments are ordered by popularity.

The dynamic \textbf{Sentence-level Comment Panel} appears when a reader selects a sentence and displays all comments anchored to that sentence. This allows readers to inspect detailed commentary while remaining at the corresponding location in the article.

Together with paragraph-level access, these panels allow readers to move among global-, paragraph-, and sentence-level commentary without leaving the primary reading context.

\subsubsection{Control and Filtering}

The top control panel allows readers to adjust how contextual commentary is presented.

Readers can switch among the Sentence-End, Between-Line, and Click-to-Show layouts and independently toggle paragraph-level comments.

Two sliders filter sentence-level comments according to like count and reply count, allowing readers to reduce low-activity content.

\subsection{Alignment Data and Training}
\label{app:alignment_details}

\subsubsection{Textual Anchor Formulation}
\label{sec:textual_anchor_formulation}

To support contextual presentation, each comment is mapped to the portion of the article it addresses. We use three levels of granularity:

\begin{itemize}
    \item \textbf{Sentence-level}: refers to one or more specific sentences.
    \item \textbf{Paragraph-level}: refers to one or more complete paragraphs.
    \item \textbf{Global-level}: refers to the article as a whole.
\end{itemize}

These anchors enable comments to be presented at corresponding levels in the interface.

\subsubsection{Dataset and Annotation}
\label{sec:data_preparation}

We collected ten long-form Zhihu articles and the top 100 comments for each, covering topics including technology, society, and history, resulting in 1,000 comment samples.

Two authors independently annotated each sample with (1) a positional level and (2) the corresponding text segment(s), such as sentence indices. Inter-annotator agreement was assessed on 20\% of the dataset, yielding Cohen's $\kappa=0.91$ for positional-level annotations.

The data were split by article to reduce leakage across related comments: eight articles (800 samples) were used for training and two articles (200 samples) for testing. Each sample was converted into an instruction--input--output format containing the indexed article text, the comment, and its structured anchor labels.

\subsubsection{Model Fine-Tuning and Inference}
\label{sec:finetuning}

We fine-tuned Qwen3-8B using LoRA with the LLaMA-Factory framework. The model was trained for three epochs on the 800 training samples with a maximum sequence length of 4,096 tokens.

At inference time, an article is segmented into indexed sentences and provided to the model together with each comment. The model outputs the positional level and textual anchor in structured JSON format. The interface then uses these outputs to place comments and corresponding visual markers at the appropriate locations.

\subsection{Prompt for Comment--Text Alignment}
\label{app:location_prompts}
The alignment module determines the textual anchor of a comment. To assist the model in navigating long-form content, we provide the article text as the context window.
\begin{PromptCode}
System Instruction:
You are an expert annotator specializing in discourse analysis. Your task is to locate the exact reference anchor of a user comment within the provided article context.
Guidelines for Positioning:
Sentence-level: Comment targets one or more specific, identifiable sentences (Indices must be integers).
Paragraph-level: Comment discusses the topic or thematic content of one or more full paragraphs.
Global-level: Comment provides a general summary or impression of the entire article.
Strict Constraints:
Output MUST be a single, valid JSON object.
The 'location_level' field MUST be one of: ["Sentence-level", "Paragraph-level", "Global-level"].
For 'Sentence-level', 'target_indices' should be a list of integers (e.g., [12, 13]).
If the anchor is ambiguous, use the most dominant context.
User Input Format:
[Context]
Indexed Article: "{article_text}"
[Comment]
User Comment: "{comment_text}"
Task: Locate the reference in the article and output in JSON.
\end{PromptCode}

\FloatBarrier
\subsection{Study Sequences}
\label{app:study_sequences}

Table~\ref{tab:study_sequences} lists the eight article--layout sequences in presentation order. Each sequence was assigned to three participants. A denotes student employment, B a public controversy, C an industrialization counterfactual, and D dynastic longevity. Each article--condition combination and each condition--order position included six participants.

\begin{table}[htbp]
\centering
\caption{Article--layout assignments in presentation order. Each sequence was assigned to three participants.}
\label{tab:study_sequences}
\begin{tabular}{ccccc}
\toprule
\textbf{Sequence} & \textbf{First} & \textbf{Second} & \textbf{Third} & \textbf{Fourth} \\
\midrule
1 & A--TE & C--SE & B--CS & D--BL \\
2 & C--CS & D--TE & A--BL & B--SE \\
3 & B--SE & A--BL & D--TE & C--CS \\
4 & D--BL & B--CS & C--SE & A--TE \\
5 & B--BL & A--CS & D--SE & C--TE \\
6 & A--SE & C--BL & B--TE & D--CS \\
7 & D--CS & B--TE & C--BL & A--SE \\
8 & C--TE & D--SE & A--CS & B--BL \\
\bottomrule
\end{tabular}
\end{table}

\begin{figure}[htbp]
  \centering
  \includegraphics[width=\linewidth]{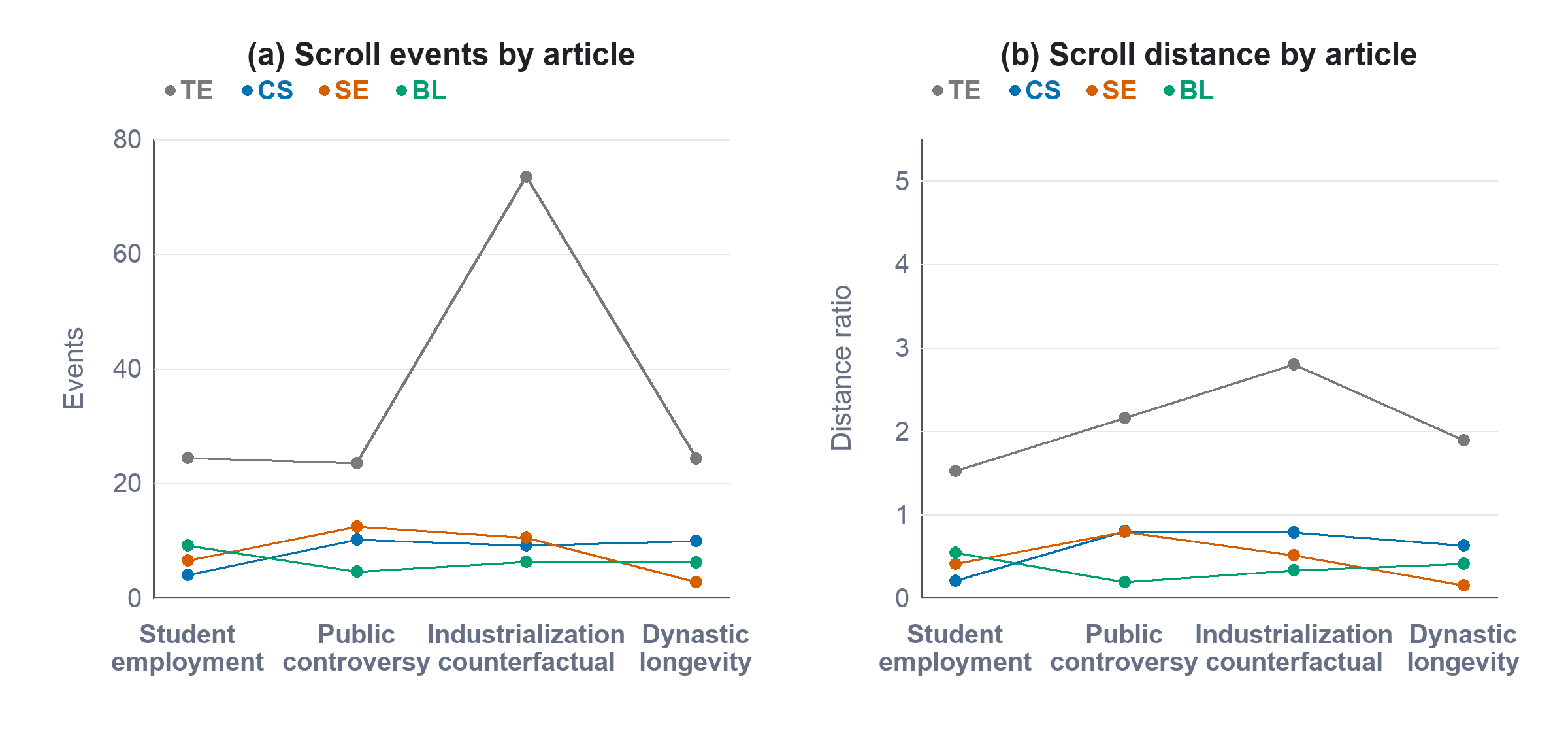}
  \caption{Exploratory navigation summaries by article topic. (a) Mean initial-reading scroll-event count. (b) Mean total distance of all upward and downward scrolling divided by the page height recorded at the end of reading. Each marker averages six participant observations for one article--condition combination. Colors identify Text-End (TE), Click-to-Show (CS), Sentence-End (SE), and Between-Line (BL); connecting lines link the condition means across article categories. Both vertical axes are linear.}
  \Description{Panels a and b show article-specific mean scroll events and normalized distance. The horizontal categories describe student employment, a public controversy, an industrialization counterfactual, and dynastic longevity. Four colored condition lines connect six-person means for each topic.}
  \label{fig:article_specific_navigation}
\end{figure}

\begin{samepage}
\subsection{Article-Specific Navigation Patterns}
\label{app:article_navigation}

The aggregate navigation pattern appeared in all four articles: TE had the highest mean scroll-event count (Fig.~\ref{fig:article_specific_navigation}a) and normalized scroll distance (Fig.~\ref{fig:article_specific_navigation}b) for every article. The ordering of CS, SE, and BL varied across topics. Each article--condition mean included six participants, with article topic treated as a categorical grouping.

\end{samepage}

\subsection{Study Material Preparation and Presentation}
\label{app:study_materials}

\subsubsection{Article Sources}
\label{app:material_sources}
We selected four Chinese long-form articles from saved Zhihu records. Articles A--D concerned student employment, a public controversy, an industrialization counterfactual, and dynastic longevity, respectively. Selection emphasized accessible topics and variation in subject matter across the reading tasks. The study retained the article bodies in these records without further editing.

\subsubsection{Comment Preparation and Rationale}
We prepared ten comments for each article through model-assisted drafting and editing. Each comment responded to a specific passage and explained the reasons for its view. The comments included agreement, disagreement, supplementary examples, and personal reflection, expressed in everyday language. These passage-linked comments provided a common basis for comparing how the four layouts supported access to commentary and its connection to the article.

Each article and its associated comments were held constant across layouts. The counterbalanced assignment distributed each article--comment set equally across the four conditions. The application loaded these fixed materials during participant sessions.

\subsubsection{Comment Anchors and Distribution}
Each comment was paired with a quotation identifying the article passage it addressed. Each article contained ten distinct anchor quotations, all uniquely located in the corresponding text. The application mapped these quotations to sentence indices for local presentation. Table~\ref{tab:study_material_characteristics} reports the article and comment lengths and the distribution of anchor locations.

Comments ranged from 75 to 128 characters. Lengths exclude whitespace and include punctuation, numbers, and Latin letters. To summarize comment locations, we divided each article into three equal parts by character count. We then counted each comment in the part where its associated quotation began.

\begin{table}[htbp]
\centering
\small
\caption{Article and comment lengths in characters, and the number of comments associated with the beginning, middle, and end of each article. Each article contained ten comments.}
\label{tab:study_material_characteristics}
\begin{tabular}{lrrccc}
\toprule
\textbf{Article} & \textbf{Article length} & \textbf{Comment length range} & \multicolumn{3}{c}{\textbf{Comment locations}} \\
\cmidrule(lr){4-6}
 & & & \textbf{Beginning} & \textbf{Middle} & \textbf{End} \\
\midrule
A & 2,704 & 86--102 & 6 & 3 & 1 \\
B & 2,459 & 75--95 & 6 & 2 & 2 \\
C & 3,434 & 109--127 & 5 & 2 & 3 \\
D & 2,226 & 106--128 & 3 & 4 & 3 \\
\bottomrule
\end{tabular}
\end{table}

\subsubsection{Comment Presentation}
\label{app:material_presentation}
SE displayed full comments after their anchor sentences. BL showed single-line previews between text segments, using an ellipsis when the text exceeded the available width; selecting a preview expanded it. CS showed the selected sentence's comments in a panel, whereas TE displayed the complete list below the article.

\end{document}